\documentclass[a4paper,fleqn,usenatbib]{mnras}

\usepackage{newtxtext,newtxmath,mathcomp}

\usepackage[T1]{fontenc}
\usepackage{ae,aecompl}

\usepackage{graphicx}	
\usepackage{amsmath}	
\usepackage{amssymb}	

\newcommand{\rev}[1]{{#1} }

\defcitealias{2017MNRAS.466.4053J}{Paper I}

\title[Observational signatures of broken discs]{Signatures of broken protoplanetary discs in scattered light and in sub-millimetre observations}

\author[S. Facchini, A. Juh\'{a}sz and G. Lodato]{Stefano Facchini$^{1}$\thanks{facchini@mpe.mpg.de}, Attila Juh\'{a}sz$^2$\thanks{juhasz@ast.cam.ac.uk} and Giuseppe Lodato$^3$\\
$^1$Max-Planck-Institut f\"ur Extraterrestrische Physik, Giessenbachstrasse 1, 85748 Garching, Germany\\
$^2$Institute of Astronomy, Madingley Road, Cambridge CB3 OHA, UK\\
$^3$Dipartimento di Fisica, Universit{\`a} degli Studi di Milano, via Celoria, 16, Milano, I-20133, Italy\\
}

\date{Accepted XXX. Received YYY; in original form ZZZ}

\pubyear{2017}

\begin{document}
\label{firstpage}
\pagerange{\pageref{firstpage}--\pageref{lastpage}}
\maketitle

\begin{abstract}
Spatially resolved observations of protoplanetary discs are revealing that their inner regions can be warped or broken from the outer disc. A few mechanisms are known to lead to such 3D structures; among them, the interaction with a stellar companion. We perform a 3D SPH simulation of a circumbinary disc misaligned by $60\degr$ with respect to the binary orbital plane. The inner disc breaks from the outer regions, precessing as a rigid body, and leading to a complex evolution. As the inner disc precesses, the misalignment angle between the inner and outer discs varies by more than $100\degr$. Different snapshots of the evolution are post-processed with a radiative transfer code, in order to produce observational diagnostics of the process. Even though the simulation was produced for the specific case of a circumbinary disc, most of the observational predictions hold for any disc hosting a precessing inner rim. Synthetic scattered light observations show strong azimuthal asymmetries, where the pattern depends strongly on the misalignment angle between inner and outer disc. The asymmetric illumination of the outer disc leads to azimuthal variations of the temperature structure, in particular in the upper layers, where the cooling time is short. These variations are reflected in asymmetric surface brightness maps of optically thick lines, as CO $J$=3-2. The kinematical information obtained from the gas lines is unique in determining the disc structure. The combination of scattered light images and (sub-)mm lines can distinguish between radial inflow and misaligned inner disc scenarios.
\end{abstract}




\section{Introduction}
\label{sec:introduction}

Scattered light observations of protoplanetary discs are revealing that many systems might host misaligned inner discs. The three dimensional (3D) structure of these systems is inferred by the peculiar pattern of shadows that are cast by the unresolved inner regions onto the outer regions of such discs. The most famous example is HD142527 \citep{2015ApJ...798L..44M}, but other cases have recently been observed, showing a similar pattern \citep[e.g.][]{2016A&A...595A.113S,2016PASJ...68L...3O,2017A&A...597A..42B, 2017ApJ...838...62L}. Other objects reveal a mild warp in the inner regions \citep[e.g.][]{2017ApJ...835..205D}, with no evidence of a radial discontinuity in the direction of the angular momentum of the disc.

Additional support to warped or broken structures in the inner regions of discs is provided by the kinematics obtained via rotational lines of abundant molecules in the (sub-)mm spectral range, in particular CO. The 1st moment map of the inner regions of a few discs deviates from simple Keplerian rotation, and is compatible with a gradient of the plane where the disc rotates \citep[e.g.][]{2014ApJ...788L..34P,2015ApJ...811...92C,2016ApJ...830L..16B,2017ApJ...840...23L,2017arXiv171000703W}. In the circumbinary GG Tau disc, there is indirect evidence of a misalignment between the outer disc and the binary orbital plane, which should also lead to a warped structure \citep{2017A&A...599A.102C}. Finally, optical and near-infrared (NIR) light curves of T Tauri stars sometimes show long-lived deep dimming events, which have been also modelled as due to a misaligned precessing inner disc \citep[e.g.][]{2013A&A...557A..77B,2013MNRAS.433.2157L,2016A&A...596A..38F,schneider_subm}. Similarly, some models of the so-called dippers \citep[e.g.][]{2010A&A...519A..88A,2014AJ....147...82C,2015AJ....149..130S,2016ApJ...816...69A} invoke a misaligned inner disc \citep[][and references therein]{2017MNRAS.470..202B} to explain the variability of about $10-50\%$ of the observed flux.

Theoretically, the approach to the possibility of misaligned inner discs has come in two different and in most cases complementary ways. The first one is to describe the (magneto-)hydrodynamical causes that might lead to the formation of broken discs, and more generally of warps. The second one consists in running radiative transfer codes on a simplified parametric model of a warped or broken disc, and then comparing synthetic images of the model with observed systems to constrain the main parameters of the 3D structure of the disc \citep[e.g.][]{2017A&A...604L..10M}. \citet{2015A&A...579A.110R} and \citet[][hereafter \citetalias{2017MNRAS.466.4053J}]{2017MNRAS.466.4053J} have started to post-process hydrodynamical simulations of mildly warped discs with radiative transfer codes, in order to produce more realistic synthetic images to be compared against observations and to provide time evolution of such structures.

Linear theory predicts that the propagation of warps in protoplanetary discs is expected to occur via bending waves \citep[][]{1983MNRAS.202.1181P,1995ApJ...438..841P}. This holds true whenever the dimensionless turbulent parameter $\alpha\lesssim H_{\rm p}/r$, where $H_{\rm p}$ is the pressure scale height of the disc \citep[see][for a recent review]{2016LNP...905...45N} and $r$ is the disc spherical radius. Both theoretical expectations \citep[e.g.][]{2014prpl.conf..411T} and constraints from observations  \citep[e.g.][]{2015ApJ...813...99F,2017ApJ...843..150F,2016A&A...592A..49T} confirm that this is indeed the case, under the assumption that turbulence is related to the kinematic viscosity $\nu=\alpha c_{\rm s} H_{\rm p}$ \citep{1973A&A....24..337S}, where $c_{\rm s}$ is the sound speed. Linearised equations of warp propagation in the bending waves regime have been derived for discs misaligned to an axisymmetric potential \citep[e.g.][]{2000ApJ...538..326L}, which can occur whenever the disc is misaligned with respect to an internal or external massive companion orbit \citep[e.g.][]{2001ApJ...560..997L,2014MNRAS.442.3700F}, or when the disc is misaligned to the magnetic dipole moment of the central star \citep[e.g.][]{2011MNRAS.412.2799F}. The steady-state solution of these equations have also been recovered by full 3D hydrodynamical simulations for different central potentials \citep{2013MNRAS.433.2142F,2015MNRAS.448.1526N}.

A detailed non linear theory of warp propagation in the bending waves regime is yet to be developed \citep[an initial study is by][]{2006MNRAS.365..977O}. However, hydrodynamical simulations have shown that when a thick disc ($H_{\rm p}/r>\alpha$) is misaligned by $\gtrsim15-30\degr$ to the symmetry plane of the potential (such as the binary orbital plane for circumbinary discs, or the plane perpendicular to the dipole moment of the stellar magnetic field), the disc can break, with (at least) an inner annulus that becomes misaligned to the outer disc \citep[e.g.][]{1996MNRAS.282..597L,2013MNRAS.433.2142F,2015MNRAS.448.1526N}. This behaviour is also reproduced in simulations of the so-called diffusive regime, i.e. when  $H_{\rm p}/r<\alpha$ \citep[e.g.][]{2012ApJ...757L..24N,2013MNRAS.434.1946N,2015MNRAS.449.1251D}, indicating that disc breaking can occur in very different astrophysical conditions. This phenomenon is known as disc tearing \citep{2012ApJ...757L..24N}, and it is the focus of this paper.

Other recent studies also show that a planet that is massive enough to carve a gap in the gas surface density of a disc can become misaligned to the outer disc by secular interaction with an external misaligned companion \citep[][]{2016ApJ...817...30L,2016MNRAS.458.4345M}, or by precessional resonances \citep{2017MNRAS.469.2834O}. In both cases, the inner disc (within the planet/companion orbital radius) might get aligned to the orbital plane of the planet, thus becoming misaligned to the outer disc. Thus, the three most promising scenarios to produce an inner disc misaligned to the outer disc are: 1) a circumbinary disc misaligned to the binary orbital plane; 2) a stellar magnetic dipole moment that is misaligned to the angular momentum vector of the outer disc; 3) a misaligned massive planet/companion tilting the circumprimary disc.

In this paper, the misaligned circumbinary case is considered, since hydrodynamical simulations are well tested for this scenario. We follow the evolution of such circumbinary case using 3D smoothed particle hydrodynamics (SPH) simulations, where the disc is expected to break due to the large initial misalignment. The simulation is used as input for radiative transfer code, in order to obtain observational diagnostics of broken discs for realistic hydrodynamical simulations. This method has already been exploited in \citetalias{2017MNRAS.466.4053J} to derive observational diagnostics of mildly warped discs, whereas in this paper we focus on the regime where the disc breaks. We stress that we focus on the circumbinary disc case, but the observational expectations obtained at different time in the disc evolution can be applied to any disc that hosts a misaligned (broken) inner disc. Only the quantitive behaviour of the secular evolution will be specific for a circumbinary disc.

\begin{figure}
\center
\includegraphics[width=\columnwidth]{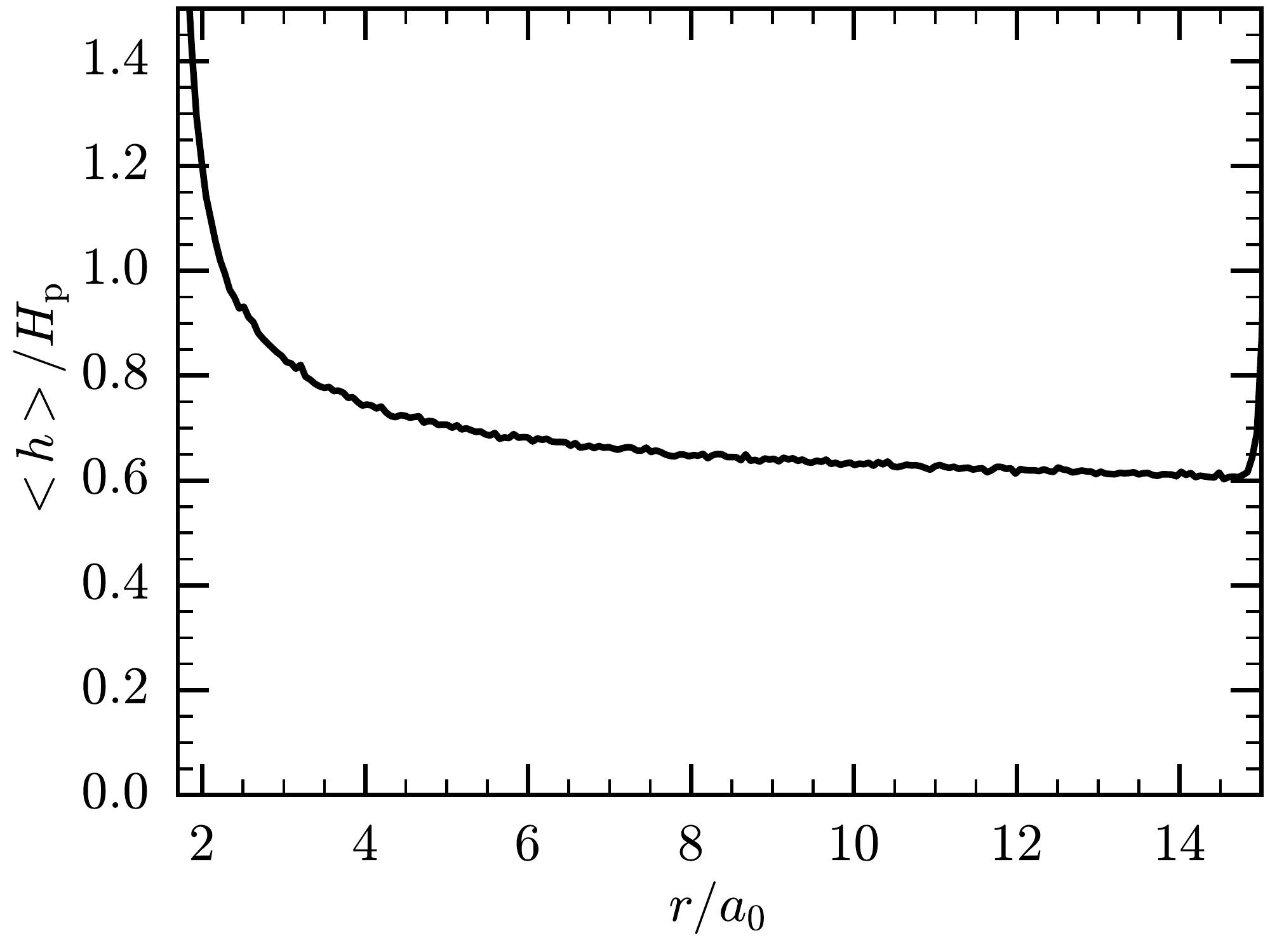}
\caption{Ratio of average smoothing length and disc scale-height at $t=0$.
}
\label{fig:h_H}
\end{figure}

\section{Hydrodynamical simulation}

\begin{figure*}
\begin{center}
\includegraphics[width=0.4\textwidth]{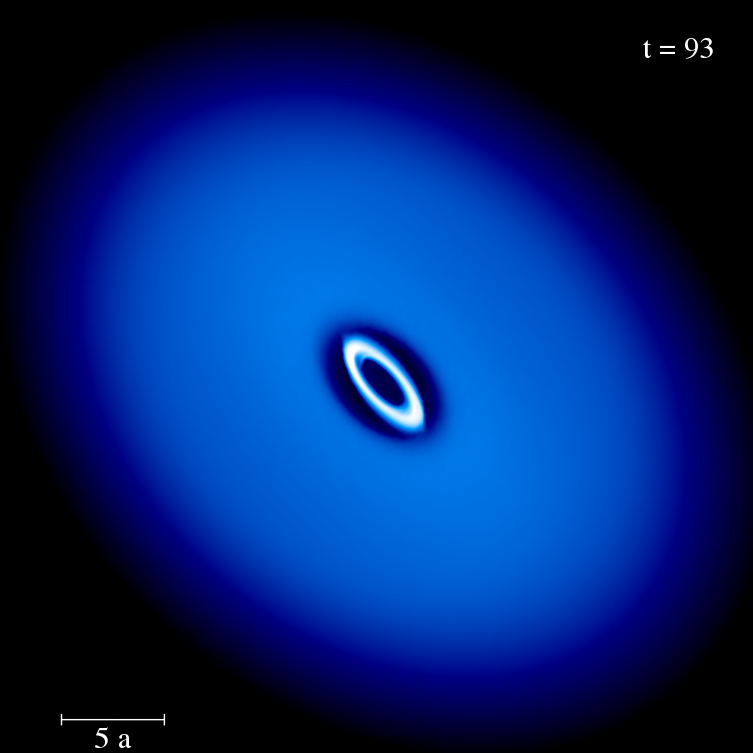}
\includegraphics[width=0.4\textwidth]{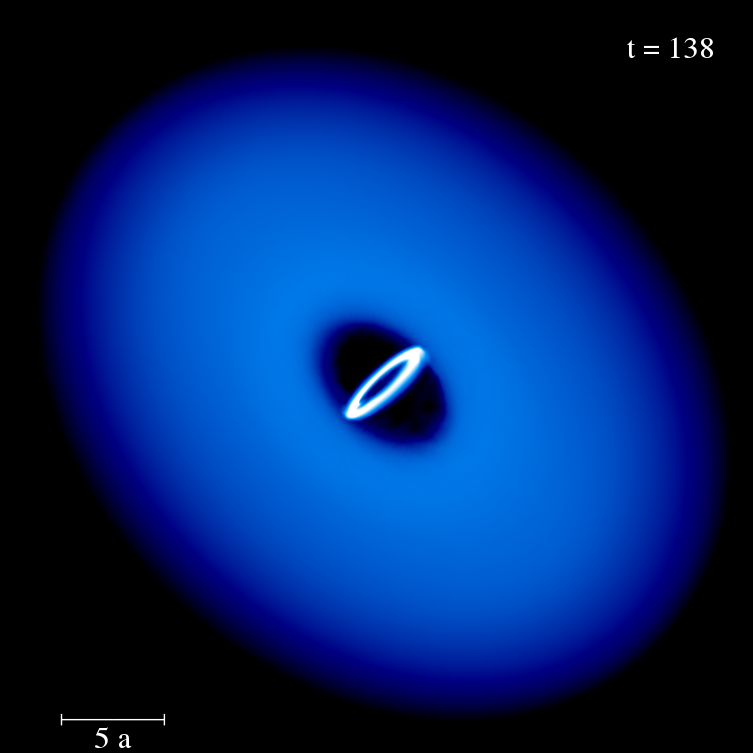}\\
\includegraphics[width=0.4\textwidth]{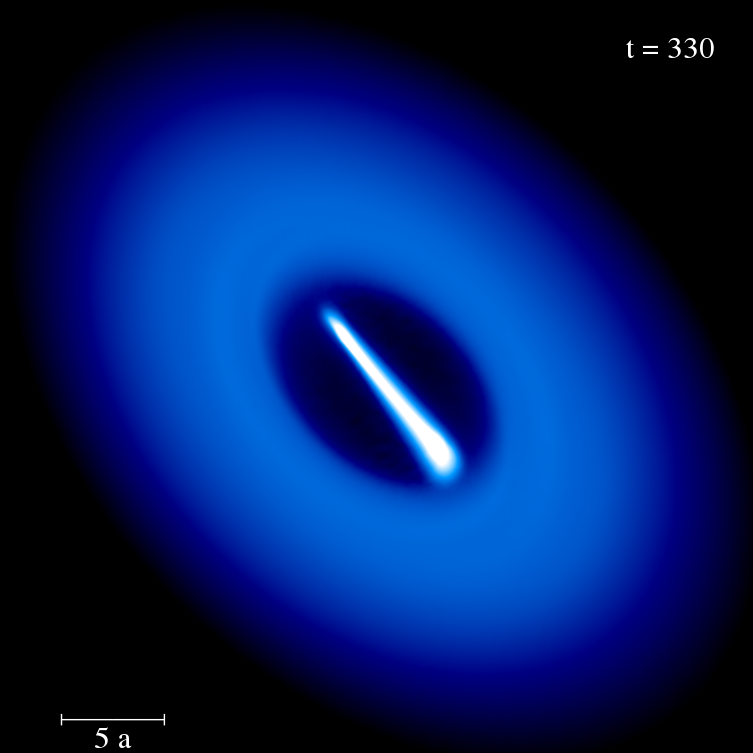}
\includegraphics[width=0.4\textwidth]{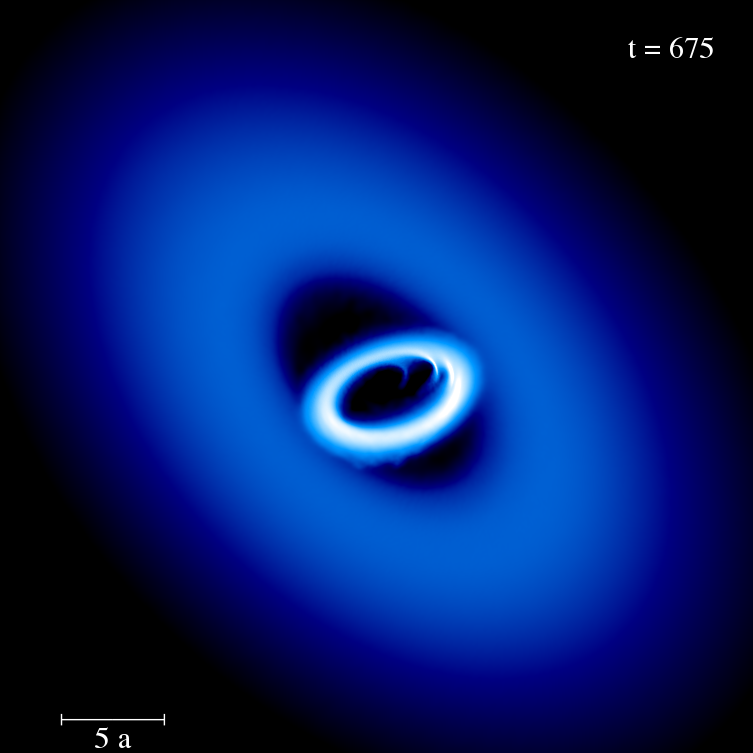}
\end{center}
\caption{Snapshots of the hydrodynamical simulation, with the vertical axis of the image being $z$, i.e. the initial direction of the binary angular momentum. The time in the top right corner is expressed in units of the initial binary period $T_{\rm b}$.}
\label{fig:evolution}
\end{figure*}

\begin{figure}
\center
\includegraphics[width=\columnwidth]{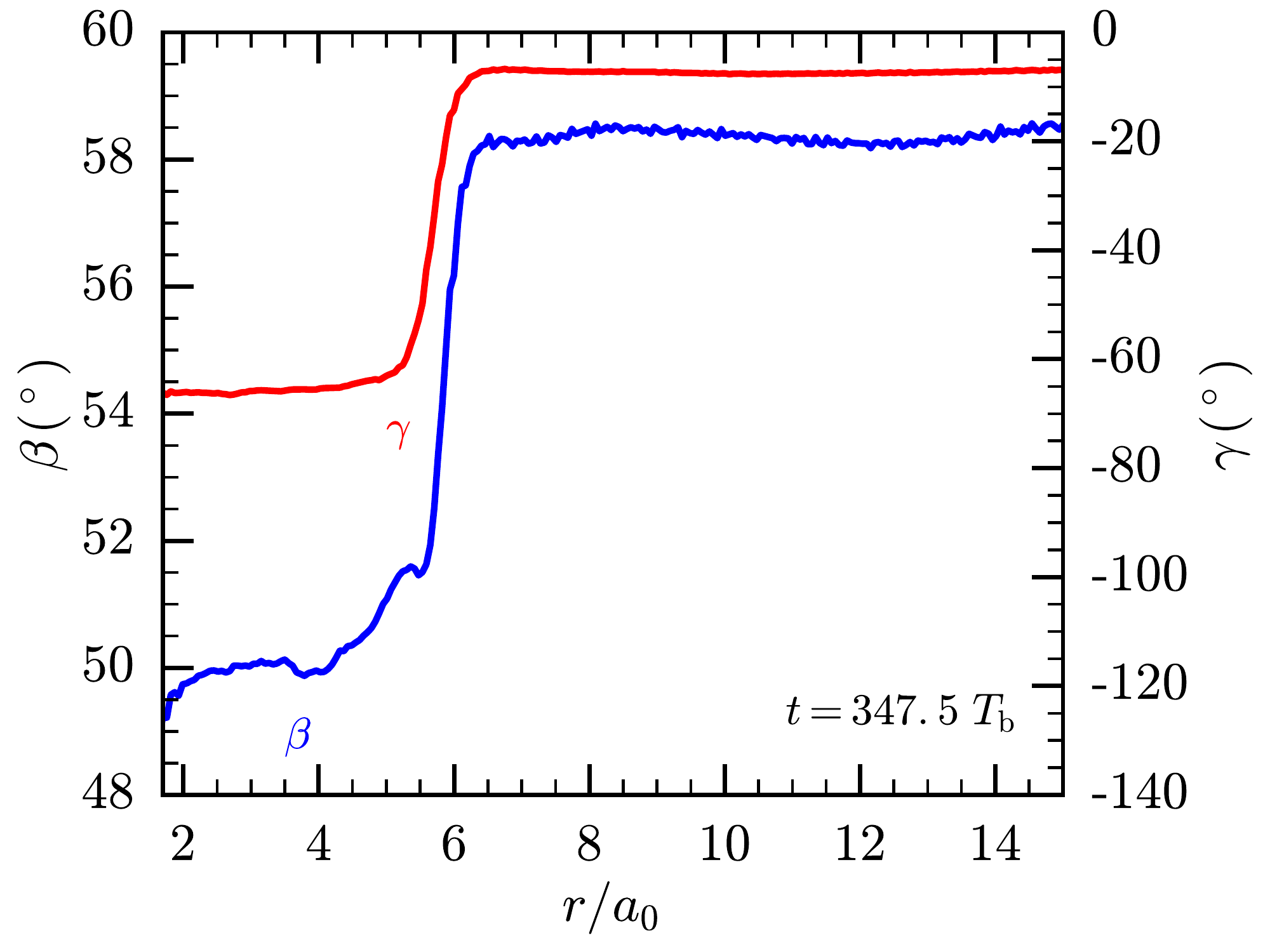}
\caption{Radial dependence of the $\beta$ and $\gamma$ angles for a snapshot at $t=347.5\,T_{\rm b}$. The distinction between the inner and outer discs is clearly visible in both angles. The outer disc is neither warped nor twisted.
}
\label{fig:beta_gamma_r}
\end{figure}

\subsection{Setup}
\label{sec:setup}

The numerical method used in this paper is very similar to the one adopted in \citetalias{2017MNRAS.466.4053J}. The 3D SPH code \textsc{phantom} \citep{2017arXiv170203930P} is used to reproduce the evolution of a circumbinary disc orbiting an equal mass binary on a circular orbit, with initial orbital separation $a_0$. The primary and secondary mass are labelled as $M_1$ and $M_2$, with the two stars being of equal mass. The initial misalignment between the disc and the binary is set to $\theta=60\degr$. The two central stars are represented by sink particles \citep{1995MNRAS.277..362B} with a sink radius of $0.45a_0$. The initial surface density of the disc is set to a power law scaling as $r^{-1}$ between $1.7a_0$ and $15a_0$, with a local isothermal equation of state with $c_{\rm s}\sim r^{-1/4}$ (and thus an implied temperature $T\sim r^{-1/2}$), which is used to set the initial vertical distribution of SPH particles. The aspect ratio $H_{\rm p}/r$ is $0.041$ at $r=1.7a_0$ \citep[i.e. at the tidal truncation radius if the binary and the disc were rotating on the same plane;][]{1994ApJ...421..651A,2015ApJ...800...96L} and scales with $r^{1/4}$. The mass of the disc is set to $1\%$ of the total mass of the central binary. The simulation uses $10^6$ SPH particles.

The moderate/high number of particles is used to maintain the effective viscosity from the artificial viscosity terms lower than the physical viscosity. In the code the artificial viscosity $\alpha^{\rm AV}$ ranges between 0.01 and 1, controlled by a \citet{1997JCoPh.136...41M} switch, and $\beta^{\rm AV}=2$ \citep[see][for more details]{2017arXiv170203930P}. The physical viscosity is implemented using the formulation by \citet{1994ApJ...431..754F}, which has been benchmarked and tested also in simulations of warped discs \citep{2010MNRAS.405.1212L,2013MNRAS.433.2142F}. The bulk viscosity is set to $0$, whereas the shear viscosity is computed as $\alpha c_{\rm s}^2/\Omega_{\rm K}$, where $\alpha=0.02$, and $\Omega_{\rm K}$ is the Keplerian orbital frequency. The shear viscosity is assumed to be isotropic\footnote{An isotropic $\alpha$ is not equivalent to an effective diffusion coefficient for the radial communication of the angular momentum. While the component of the angular momentum that is parallel to the local orbital plane is radially communicated with an effective viscosity $\propto1/\alpha$, the component  perpendicular to the local orbital plane is communicated with an effective viscosity $\propto \alpha$ \citep{1983MNRAS.202.1181P,1992MNRAS.258..811P,2007MNRAS.381.1287L,2015MNRAS.450.2459N}.}. Note that \citet{2016MNRAS.455L..62N} have shown that warped disc models with isotropic $\alpha$ and turbulence driven by magneto rotational instability \citep{1991ApJ...376..214B} lead to the same warp evolution.

The effective viscosity (parametrized as $\alpha_{\rm eff}$) due to artificial viscosity can be roughly estimated as \citep[e.g.][]{2010MNRAS.405.1212L}:
\begin{equation}
\label{eq:alpha}
\alpha_{\rm eff} \approx \frac{1}{10}\frac{<h>}{H_{\rm p}}\alpha^{\rm AV},
\end{equation}
where $<h>$ is the average smoothing length at a given radius $r$. The radial dependence of  $<h>/H_{\rm p}$ at the initial condition of the simulation is shown in \autoref{fig:h_H}. Within the bulk of the disc and away from shocks, where $\alpha^{\rm AV} \sim 0.01$, the physical viscosity is much higher than the effective viscosity caused by the artificial one. In regions that are affected by shocks, where $\alpha^{\rm AV}\gtrsim 0.1$, the effective viscosity can give a non negligible contribution to the radial transport of material.

The evolution of the binary-disc system is computed for $1400$ binary orbits, where the period of the binary at $t=0$ is expressed as $T_{\rm b}$ throughout the paper.

In order to describe quantitatively the evolution of the misalignment angles in different parts of the disc, we define a reference frame and a few useful angles. The plane where the binary orbits at $t=0$ is the $xy$-plane. The disc is initially misaligned along the $y$-axis. At every radius $r$ of the disc, the inclination can be defined through the specific angular momentum vector ${\bf l}(r) = (l_x,l_y,l_z) = (\cos\gamma(r)\sin\beta(r),\sin\gamma(r)\sin\beta(r),\cos\beta(r))$ \citep[e.g.][]{1996MNRAS.281..357P}. The tilt angle $\beta(r)$ defines the angle between the direction of the specific angular momentum of the disc and the positive $z$-axis. The twist angle $\gamma(r)$ describes the azimuthal angle of the specific angular momentum with respect to an arbitrary axis on the $xy$-plane. Usually these angles are defined with respect to the specific angular momentum of the binary. However, since the binary inclination does mildly evolve during the whole simulation, we have defined these quantities with respect to a plane fixed in time.

\subsection{Results}

In a few binary orbits, the outer disc expands outwards, since the initial condition implies a pressure discontinuity at the initial outer edge. More importantly, after a few binary orbits the disc breaks in two separate annuli (\autoref{fig:evolution}), where the inner disc starts to precess independently as a rigid body, and the outer disc is only mildly warped (\autoref{fig:beta_gamma_r}). The radius at which the disc initially breaks is at $\sim3a_0$. This breaking radius is in broad agreement with the theoretical expectations by \citet[][see their Appendix]{2013MNRAS.434.1946N}. Within the assumption of an inviscid disc, they predict that in the bending waves regime, the disc should break when the external torque is larger than the internal torque generated by pressure forces. For a circumbinary case, this requirement translates into \citep[equation A3 in][]{2013MNRAS.434.1946N}:

\begin{equation}
\label{eq:bend_wave}
r_{\rm break}\lesssim \left( \frac{3}{4} \mu_2 |\sin{2\theta}| \frac{r}{H_{\rm p}} \right)^{1/2}a,
\end{equation}
where $\mu_i=M_i/(M_1+M_2)$ (equalling $0.5$ in this work for both $i=1$, $2$), and $a$ is the binary separation at time $t$. Implementing the parameters of the simulations presented here, with $H_{\rm p}/r$ evaluated at the inner edge, we obtain that $r_{\rm break}\lesssim 2.8a_0$, very close to the $3a_0$ value reported above. Note that we do expect $r_{\rm break}$ to be close to the inner edge of the disc, since the external torque generated by the binary declines steeply with radius, with a power-law form $\propto r^{-7/2}$ \citep[e.g.][]{2011MNRAS.417L..66N,2013MNRAS.433.2142F}. 

\begin{figure}
\center
\includegraphics[width=0.4\textwidth]{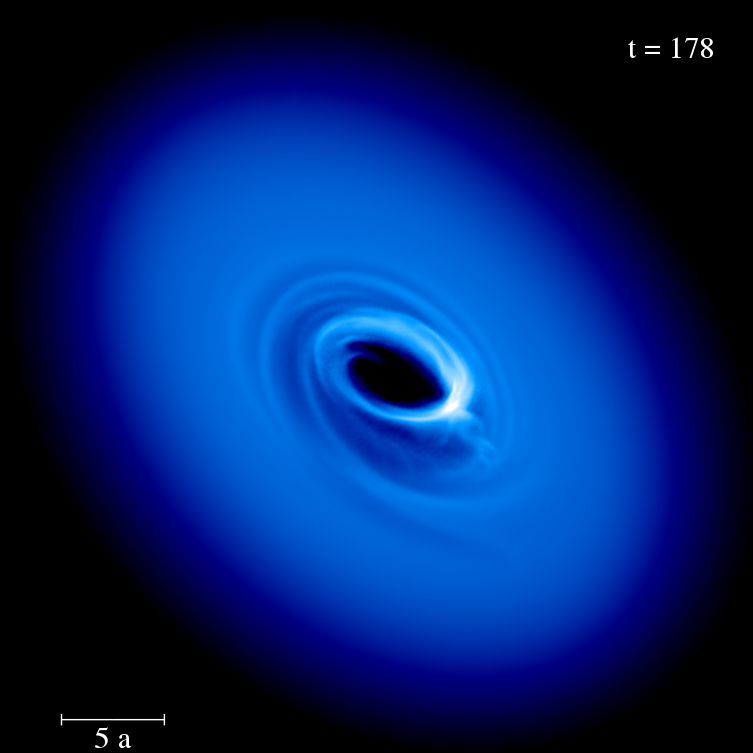}
\caption{Snapshot of the hydrodynamical simulations. When the misalignment between inner and outer disc has become high enough, the material grazing the connection region can shock and quickly accrete onto the central stars.}
\label{fig:accretion}
\end{figure}

After $\sim1$ precession period, at $t\sim175\,T_{\rm b}$, a combination of pressure gradients from the outer disc, and a high effective viscosity caused by the very low gas density and $\alpha^{\rm AV}$ being $\sim0.1$, has filled in the small gap between the inner and the outer disc. At this time, the misalignment between the inner and outer disc is also high enough that the relative velocities of the gas parcels of the inner and outer regions at radii close to $r_{\rm break}$ are supersonic. Thus, the gas undergoes a shock, with consequent loss of kinetic energy and cancellation of angular momentum (see \autoref{fig:accretion}), leading to a temporary disruption of the inner disc and an enhancement in the accretion rate \citep[as seen in other simulations, e.g.][]{2012MNRAS.422.2547N,2012ApJ...757L..24N,2013MNRAS.434.1946N,2015MNRAS.448.1526N,2015MNRAS.449.1251D}.

Mass flux from the outer to the inner regions leads to the formation of a new inner disc (visible in the two lower panels of \autoref{fig:evolution}), where the new $r_{\rm break}\sim5.5a_0$ is larger than the initial one. This is due to viscous internal torques becoming comparable to internal pressure torques. In fact, after the accretion event, the inner disc has a lower surface density, leading to higher viscous torques (a lower density implies higher numerical viscosity in SPH, see \autoref{eq:alpha}). In the extreme case of viscous torques dominating the warp evolution, \citet{2013MNRAS.434.1946N} derived the following constraint on $r_{\rm break}$:

\begin{equation}
\label{eq:viscous}
r_{\rm break}\lesssim 50 \mu_2^{1/2} |\sin{2\theta}|^{1/2} \left( \frac{H_{\rm p}/r}{0.1} \right)^{-1/2} \left( \frac{\alpha}{10^{-3}} \right)^{-1/2} a.
\end{equation}
For the parameters used in the simulation, and using the physical value of $\alpha$ implemented in the simulation, this implies $r_{\rm break}\lesssim11.5a_0$. After the disruption of the first inner disc, the breaking radius is halfway between what is predicted by the bending waves regime (\autoref{eq:bend_wave}), and by the diffusive regime (\autoref{eq:viscous}). A similar result has been found by \citet{2015MNRAS.448.1526N} for a different external torque.

For simplicity, in this simulation, we initialised the binary on a circular orbit. However, some level of eccentricity can easily be attained in astrophysical systems. A famous example is the circumbinary disc-hosting system GG Tau, where the best fit to astrometric measurements of the central stars indicates an eccentric orbit \citep{2011A&A...530A.126K}. Even a small eccentricity of the central binary can induce a particular evolution of the circumbinary disc, as polar alignment \citep{2015MNRAS.449...65A,2017ApJ...835L..28M,2017arXiv170607823Z}, but due to our initial condition we do not observe such behaviour.

Throughout the whole simulation, the binary orbital parameters do not vary significantly. The binary separation varies by a maximum of $1.5\%$, and the eccentricity increases to a maximum of $\sim2\times10^{-3}$ (see \autoref{fig:binary}). The inclination of the orbital plane evolves somehow more significantly, due to the back-reaction of the disc onto the binary itself, and the accreted material. The orbital plane tilts by a maximum of $\sim3.7\degr$ throughout the simulation.

In order to quantitatively estimate the evolution of the inclination of the inner and outer disc, we follow the method described in section $3.2.6$ of \citet{2010MNRAS.405.1212L}. We thus compute the specific angular momentum vector averaging over the gas particles contained in 500 thin concentric spherical shells between $0.75a_0$ and $30a_0$. To trace the inner disc, we consider the evolution at $r=2a_0$, i.e. at smaller radii than the breaking radius throughout the whole simulation. In order to trace the outer disc, we select $r=14a_0$, which is always well beyond the breaking radius. \autoref{fig:beta_gamma} shows the evolution of the inclination and precession of the inner and outer discs through the two quantities $\beta$ and $\cos\gamma$. The latter tracks the cosine of the angle between the projection of the specific angular momentum and the $x$-axis, i.e. showing the precession of the disc around the $z$-axis. A period in $\cos\gamma$ coincides with a precession period. Both quantities show that the outer disc does not have the time to evolve significantly: the misalignment angle between the outer disc and the $xy$-plane stays constant within $5\degr$; the outer disc rigidly precesses around the $z$-axis by $\sim30\degr$ throughout the whole simulation. The inner disc shows more interesting evolution, as already apparent in \autoref{fig:evolution}. Before $t\sim175\,T_{\rm b}$, the misalignment between the inner disc and the $xy$-plane (i.e. between the inner disc and the binary, since the inclination of the binary itself does not evolve significantly) decreases with time, due to both the inclination damping that is expected from viscous dissipation \citep[e.g.][]{2000MNRAS.317..773B,2013MNRAS.433.2157L,2016MNRAS.455.1946F}, and to the interaction with the outer disc. In the meantime, the inner disc completes an entire precession around the $z$-axis (see right panel of \autoref{fig:beta_gamma}). As already mentioned, by $t\sim175\,T_{\rm b}$ the outer disc has spread inwards, and most of the inner disc is accreted due to angular momentum cancellation. This phenomenon is at the origin of the ``noise'' between $t\sim175\,T_{\rm b}$ and $t\sim250\,T_{\rm b}$ in \autoref{fig:beta_gamma}. The inner disc is then re-formed by material inflowing from the outer disc, thus with $\beta\sim60\degr$. Thus, the misalignment of the inner disc increases between $t\sim175\,T_{\rm b}$ and $t\sim\,300\,T_{\rm b}$. As the inner disc is re-formed, the evolution is similar to the initial one. However, the second inner disc has a larger $r_{\rm break}$, thus its precession period is much longer than in the initial phase. In fact it is possible to demonstrate that the precession period of the inner disc scales with $\sim r_{\rm break}^{3/2}$ \citep[from equation 9 of][]{2013MNRAS.433.2157L}. To be more quantitative, we can compare the expected precessional period of the inner disc when $r_{\rm break}\sim3a_0$ with the one obtained from the simulation. Re-adapting equation 12 of \citet{2013MNRAS.433.2157L} to the notation used in this paper, we have that:

\begin{figure}
\center
\includegraphics[width=\columnwidth]{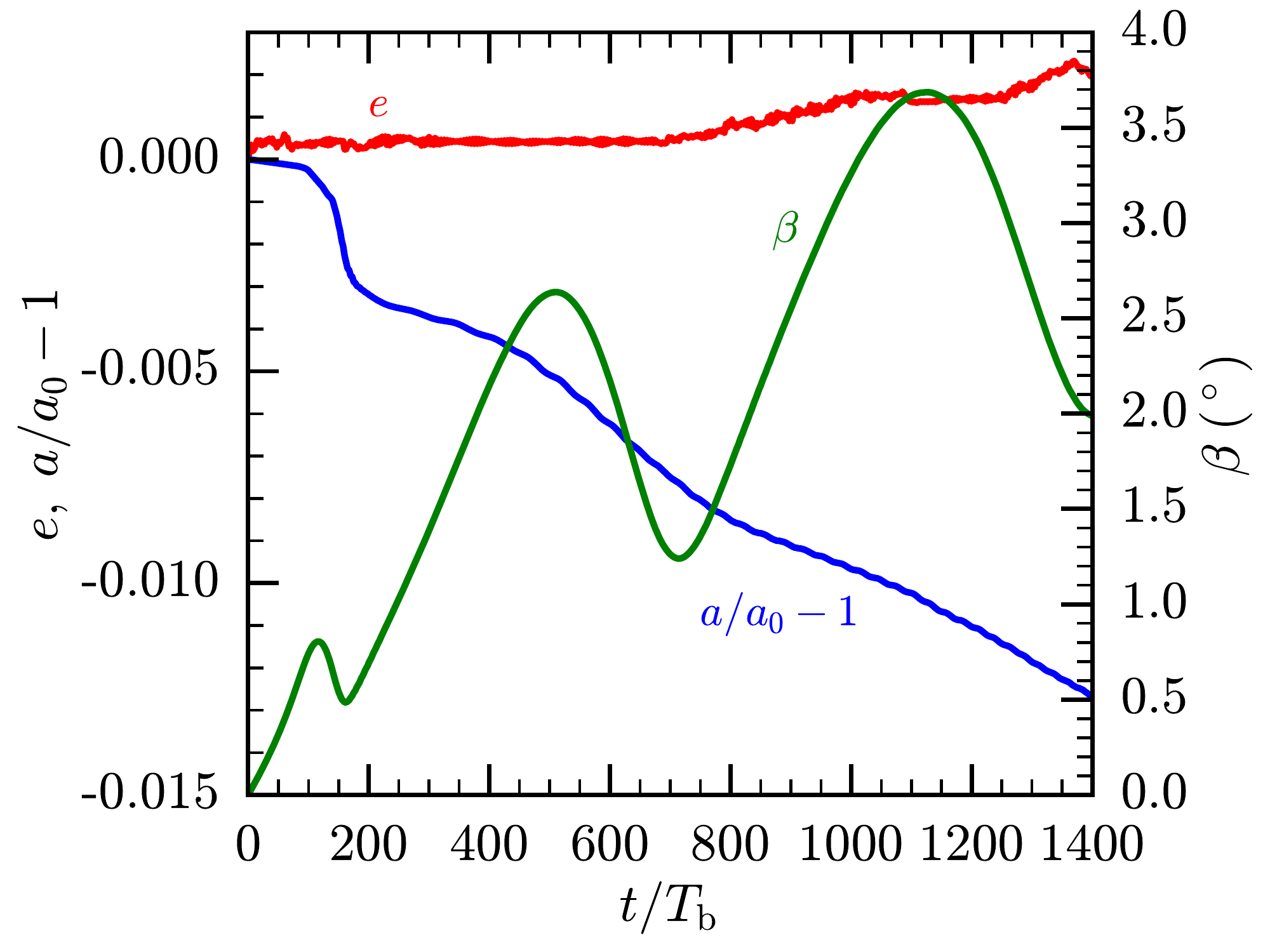}
\caption{Evolution of eccentricity, orbital separation and inclination with respect to the $xy$-plane of the central binary. Both eccentricity and orbital separation vary by a negligible amount. The inclination, expressed with the angle $\beta$, increases to a maximum value of $\sim3.7\degr$ during the whole simulation.}
\label{fig:binary}
\end{figure}

\begin{figure*}
\begin{center}
\includegraphics[width=\columnwidth]{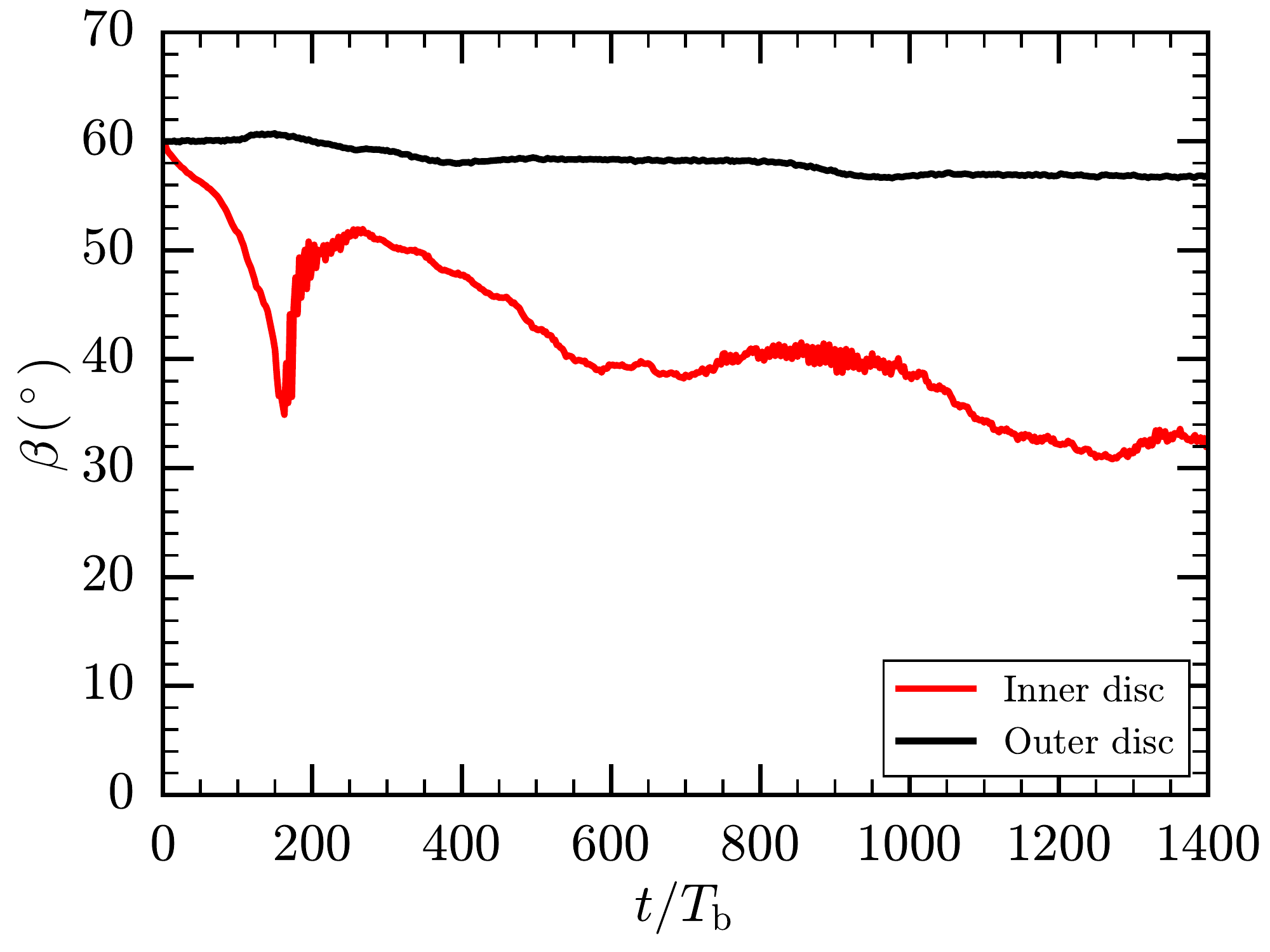}
\includegraphics[width=\columnwidth]{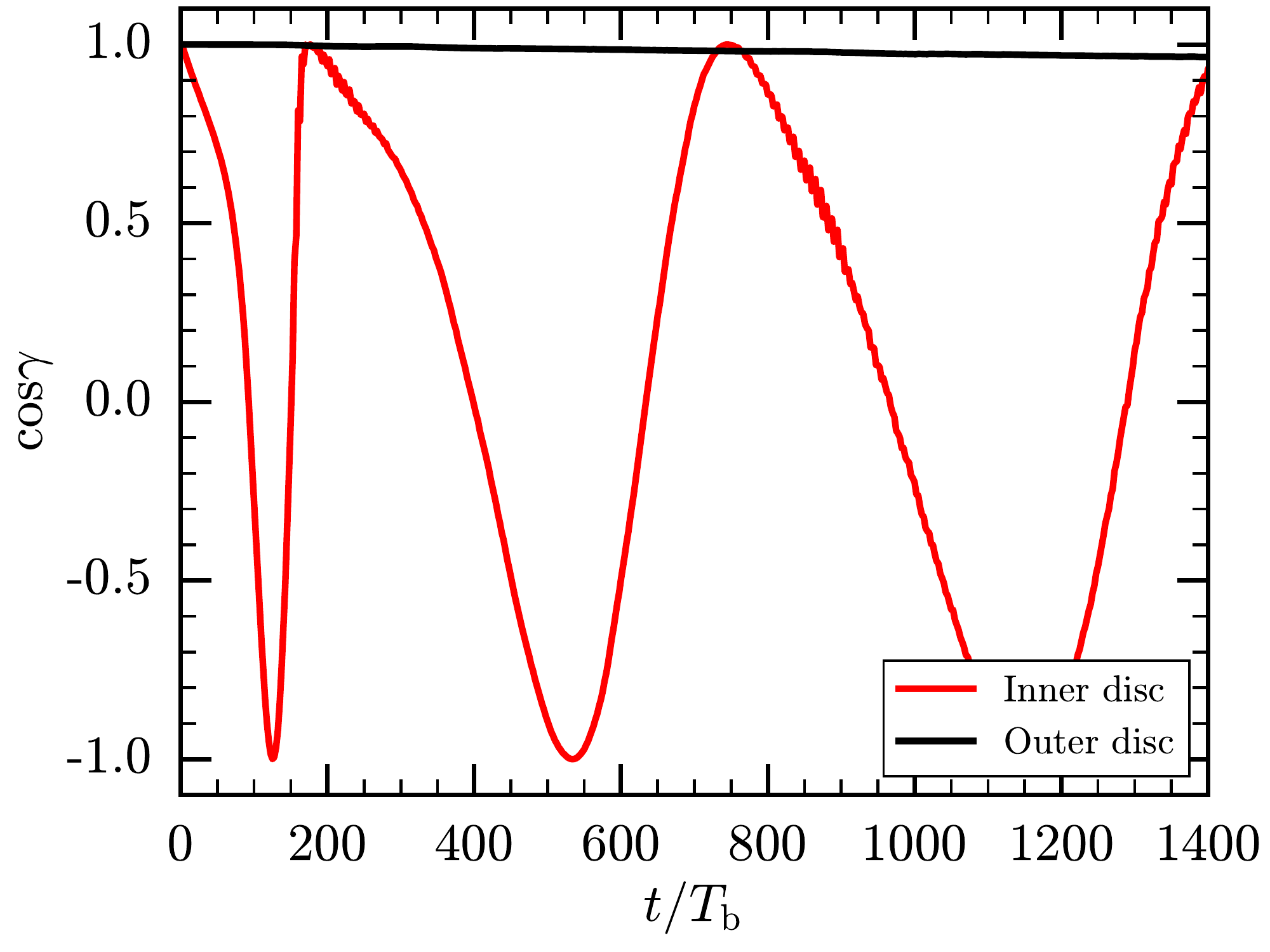}
\end{center}
\caption{Left panel: evolution of the inclination angle of the inner and outer discs, where $\beta$ is the misalignment angle of the specific angular momentum to the $z$-axis. Right panel: evolution of the cosine of the angle $\gamma$, which tracks the evolution of the rotation of the angular momentum unit vector around the $z$-axis. One period in $\cos\gamma$ indicates a whole precession. In both panels, the inner disc is traced at $r=2a_0$, whereas the outer disc is traced at $r=14a_0$ (see main text). }
\label{fig:beta_gamma}
\end{figure*}

\begin{figure}
\center
\includegraphics[width=\columnwidth]{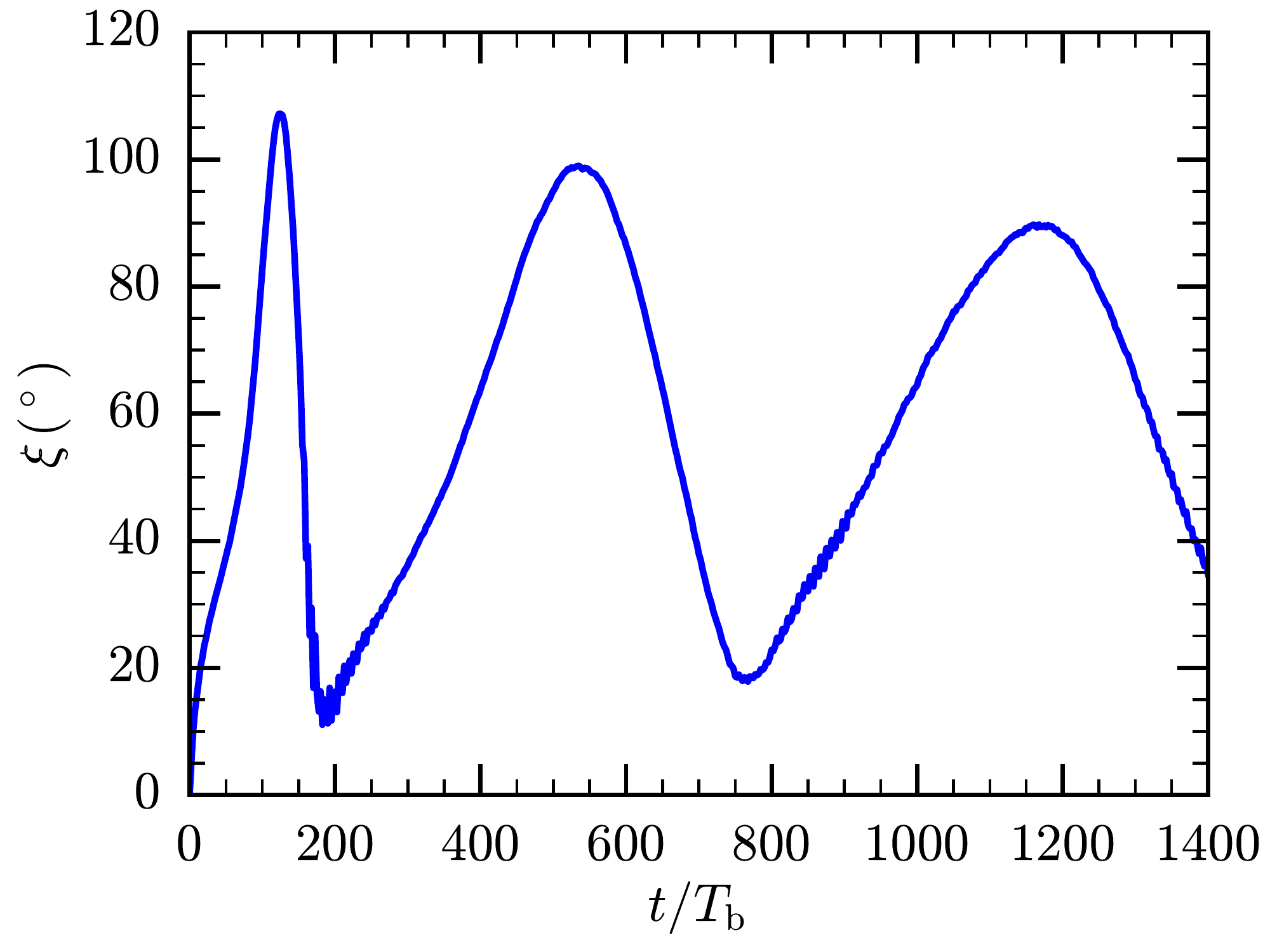}
\caption{Evolution of the misalignment angle between the inner and outer discs. The evolution is mostly set by the initial misalignment of the disc to the binary orbit, and by the precession of the inner disc.}
\label{fig:xi}
\end{figure}

\begin{equation}
T_{\rm p} = \frac{4}{3\mu_1\mu_2\cos{\bar{\beta}_{\rm in}}}\left(\frac{p+1}{5/2-p}\right) \left(\frac{r_{\rm in}}{a}\right)^{7/2}  \frac{(r_{\rm break}/r_{\rm in})^{5/2-p}-1}{1-(r_{\rm break}/r_{\rm in})^{-(p+1)}} \,T_{\rm b},
\end{equation}
where $-p$ is the power-law index of the surface density profile, defined as $-1$ in \autoref{sec:setup}, and $\bar{\beta}_{\rm in}=50\degr$ is the average inclination of the inner disc during its first precession period. The disc inner radius is $\sim1.5a$ (obtained from the simulation), where the tidal truncation radius is smaller than the canonical $1.7a$ predicted by \citet{1994ApJ...421..651A}, since the disc is not coplanar with the central binary and is thus truncated at smaller radii \citep{2015ApJ...800...96L}. Implementing the numbers used in the simulation, with $a=a_0$ since the evolution of the binary semi-major axis is negligible, we obtain that $T_{\rm p}\approx 110\,T_{\rm b}$. This is in line with the precession period observed in the simulation before the inner disc is disrupted, with $T_{\rm p}\approx 116\,T_{\rm b}$ (see right panel of \autoref{fig:beta_gamma}).

\begin{figure*}
\begin{center}
\includegraphics[width=\textwidth]{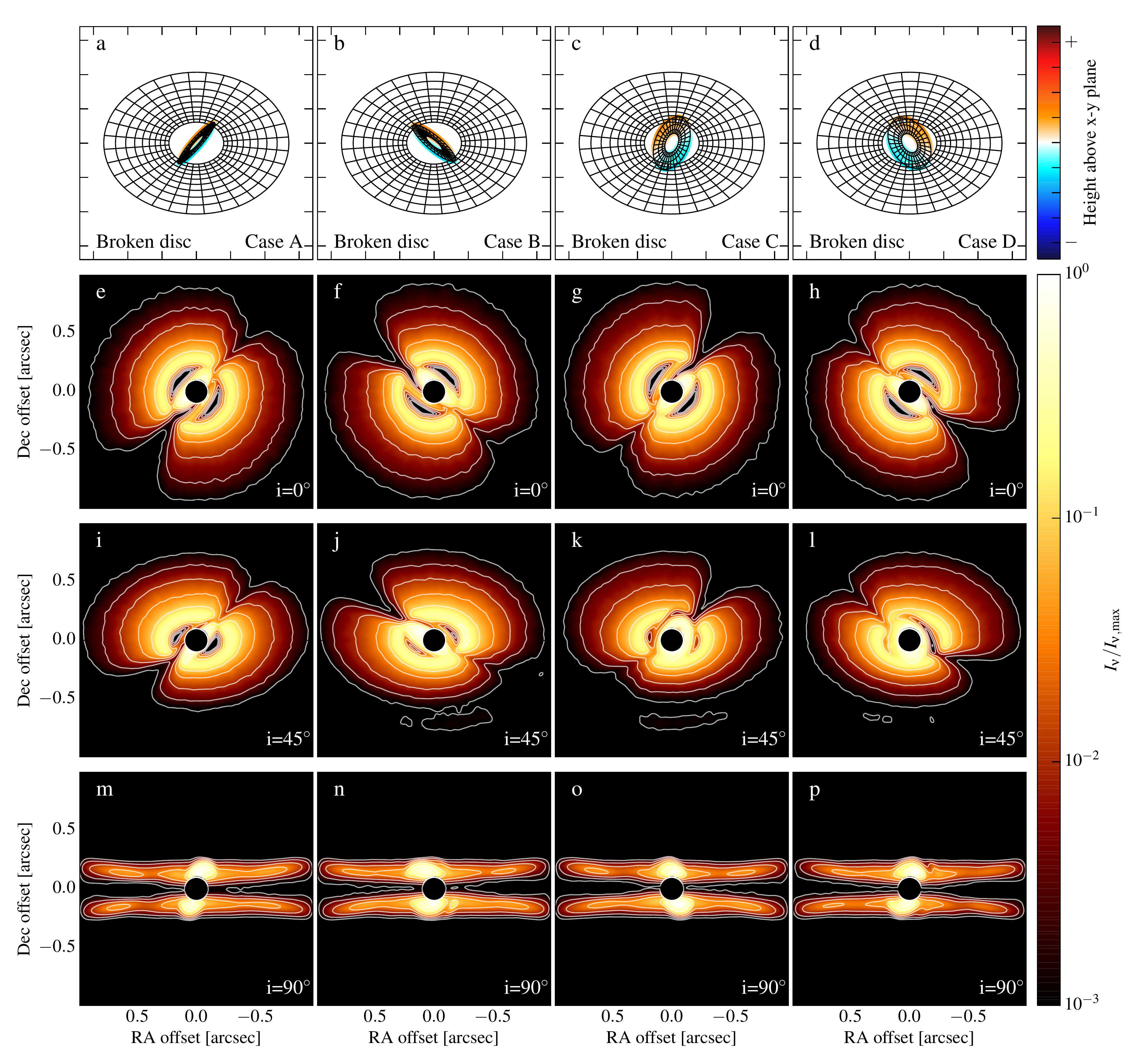}
\end{center}
\caption{Synthetic scattered light observations at $1.65\,\mu$m ($H$-band) of the hydro model at $t=430\,T_{\rm b}$, when the misalignment angle between inner and outer disc is $\xi\sim74\degr$.  The upper four panels (a-d) show a schematic of the disc 3D structure, where the misalignment angle between inner and outer disc is the one obtained for this particular snapshot of the hydrodynamical simulation, but radial distances are not on scale. From left to right: azimuth angle varies by $90\degr$ between each panel. From second row to bottom: the inclination angle varies between $0\degr$ and $90\degr$. The misaligned inner disc clearly casts an almost symmetric shadow onto the outer disc.
}
\label{fig:scat_1}
\end{figure*}

\begin{figure*}
\begin{center}
\includegraphics[width=\textwidth]{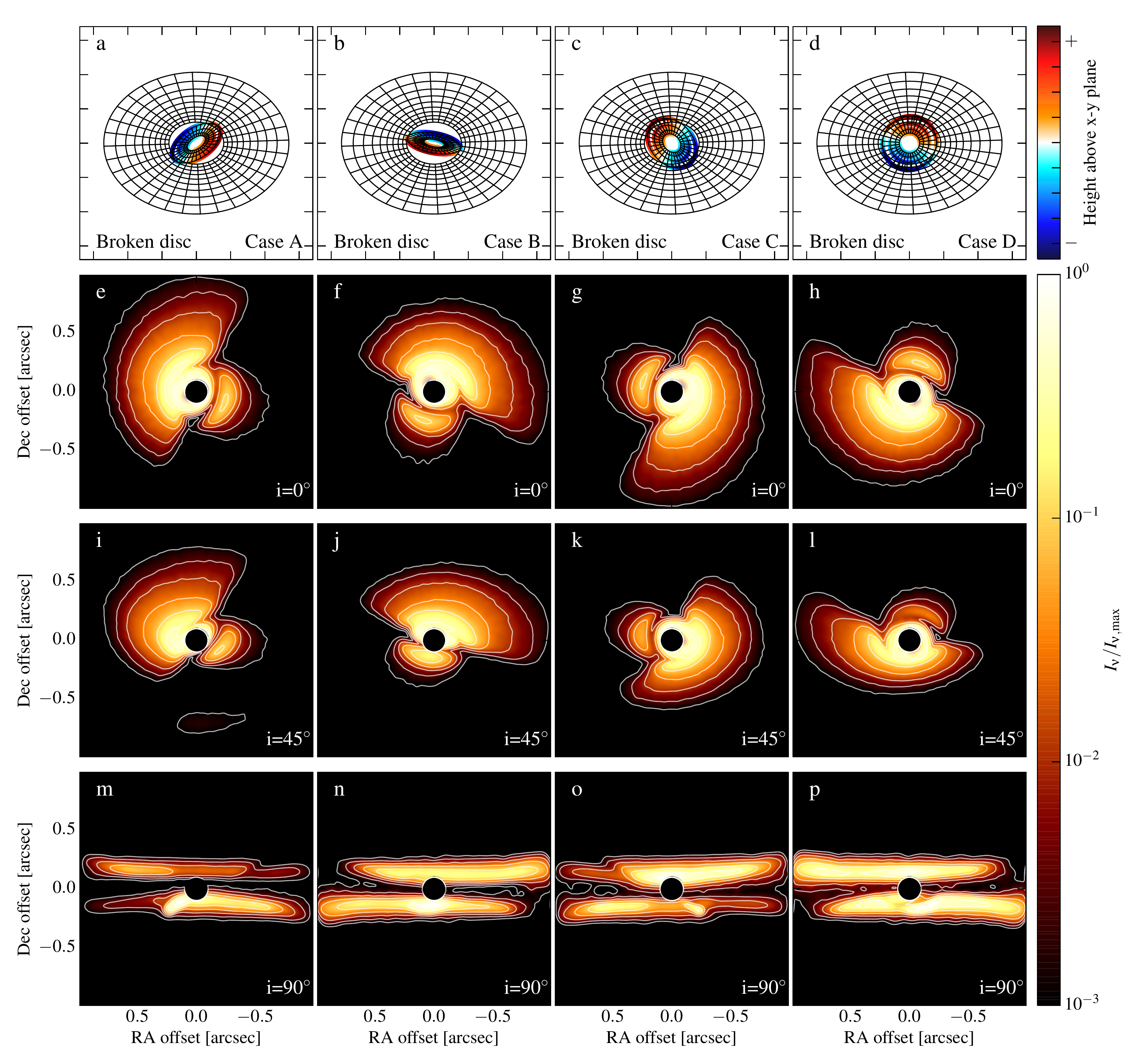}
\end{center}
\caption{Synthetic scattered light observations at $1.65\,\mu$m ($H$-band) of the hydro model at $t=270\,T_{\rm b}$, when the misalignment angle between inner and outer disc is $\xi\sim30\degr$. As in \autoref{fig:scat_1} the upper four panels (a-d) show a schematic of the disc 3D structure, where the misalignment angle between inner and outer disc is the one obtained for this particular snapshot of the hydrodynamical simulation. The panels are as in \autoref{fig:scat_1}. The pattern of the shadow is very different from the case shown in \autoref{fig:scat_1}, due to the much lower value of the $\xi$ angle in this case.}
\label{fig:scat_2}
\end{figure*}

All the angles defined so far are expressed with respect to the $xy$-plane. However, the most interesting quantity that can be constrained from observations is the misalignment angle between the inner and the outer disc. We define this angle $\xi$ as the angle between the angular momentum unit vectors of the inner and of the outer disc. The evolution of the angle $\xi$ is reported in \autoref{fig:xi}. The most important effect setting the misalignment between the two regions of the disc is the precession of the inner disc. This is apparent comparing \autoref{fig:xi} with the right panel of \autoref{fig:beta_gamma}. The peak of the misalignment between the two regions of the disc is at $t\sim125\,T_{\rm b}$, i.e. when the inner and outer discs are in antiphase. At this moment, the misalignment is equal to the sum of the two angles $\beta$.  The misalignment can be as high as $110\degr$, and as the inner disc precesses, any misalignment between this maximum value and the difference of the two $\beta$ angles can be obtained, depending on the phase of the precession.

\section{Radiative transfer}
In order to make predictions for observations we used the 3D radiative transfer code  \textsc{radmc-3d}\footnote{\url{http://www.ita.uni-heidelberg.de/~dullemond/software/radmc-3d/}}. Since \textsc{radmc-3d} is a grid-based Monte-Carlo radiative transfer code, our first step towards calculating
observables was to remap the density structure in the hydrodynamic simulations to that of the grid used in \textsc{radmc-3d}.
To do so we scaled the hydrodynamic calculations such that the  initial orbital separation $a_0$ was assumed to be 5\,AU and we assumed stellar parameters  resembling those of Herbig stars, M$_\star$=2\,M$_\odot$, R$_\star$=2.5\,R$_\odot$, T$_{\rm eff}$=9500\,K, while the disc mass within 100\,AU was assumed to be M$_{\rm disc}$=0.01M$_\odot$. In the radiative transfer calculations we used a 3D spherical mesh with N$_{\rm r}=224$, N$_{\rm \theta}=202$, N$_{\rm \phi}=200$ grid cells in the radial, poloidal and azimuthal directions, respectively. The radial grid extends between 7.5\,AU and 100\,AU and we used a logarithmic spacing for the grid cells. For the poloidal and azimuthal directions we used equidistant 
cell spacings. We applied SPH interpolation with the standard cubic spline kernel to remap the density structure in the hydrodynamic simulations
to the cell centers of the 3D spherical mesh. 

\begin{figure*}
\begin{center}
\includegraphics[width=\columnwidth]{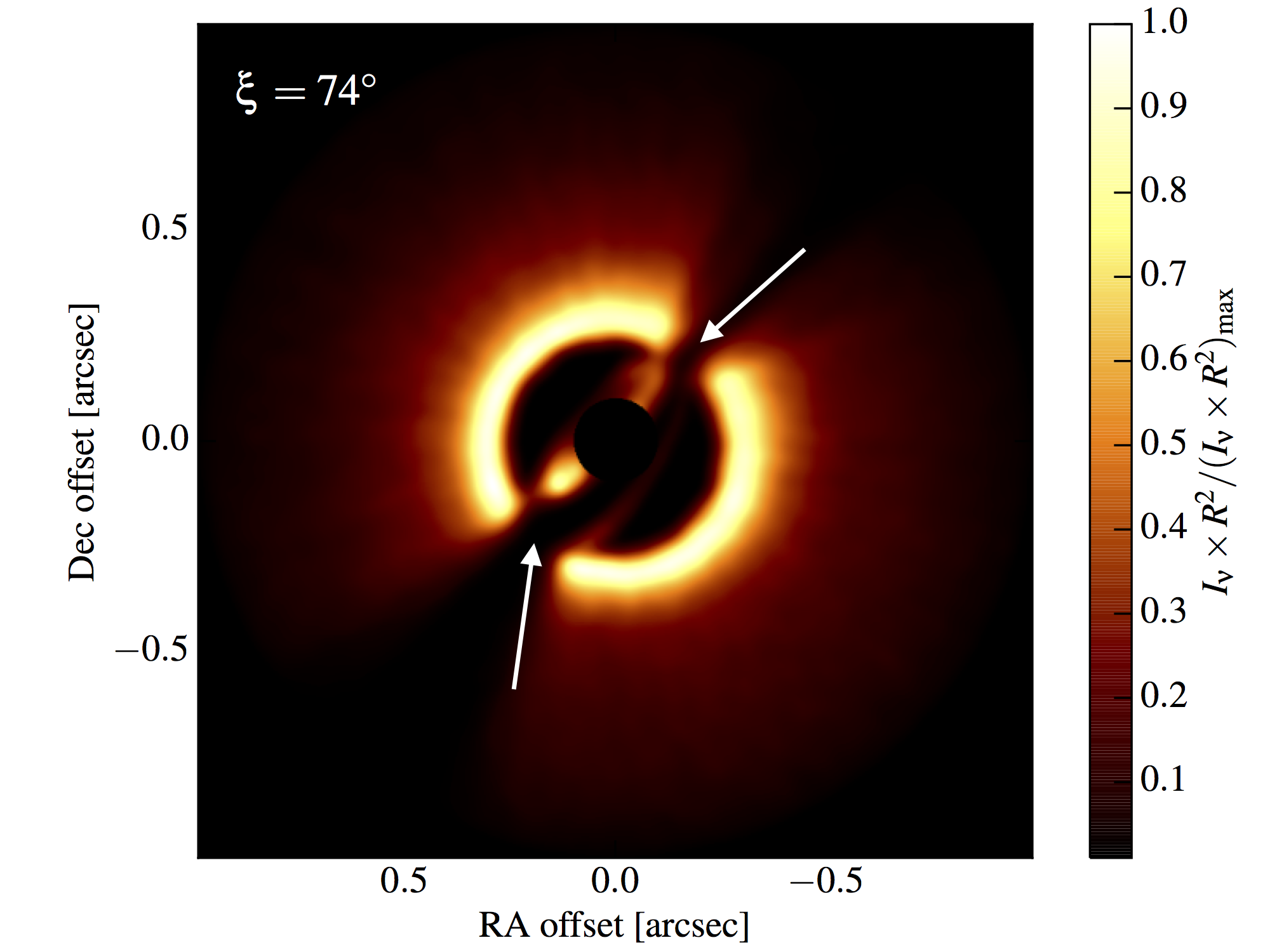}
\includegraphics[width=\columnwidth]{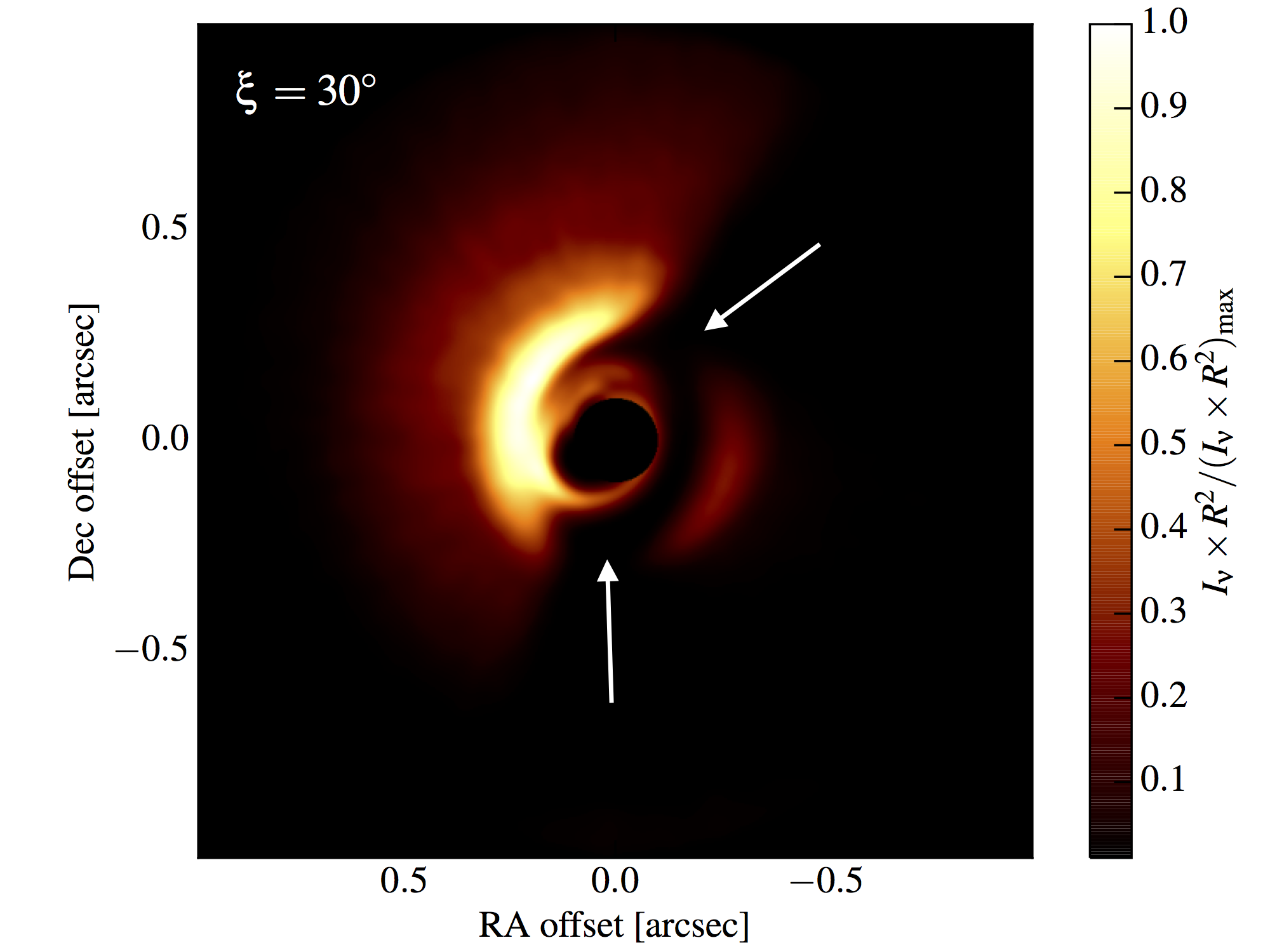}
\end{center}
\caption{Synthetic observation of scattered light emission of panel h of \autoref{fig:scat_1} (left panel) and \autoref{fig:scat_2} (right panel), when the outer disc is seen face-on. The emission has been multiplied by $R^2$ (with $R$ being the radial coordinate in the plane of the sky), to compensate for the geometric dilution of the flux of the central stars. The difference in the emission of the two cases is due to the different misalignment angles between inner and outer disc: $\sim74\degr$ in the left panel, and $\sim30\degr$ in the right panel. The two shadows in the inner region of the outer disc are highlighted by the white arrows.
}
\label{fig:scat_comp}
\end{figure*}

Dust particles in the disc had a grain size distribution between 0.1\,$\mu$m and 1\,mm with a power exponent of -3.5 \citep[see review by][and references therein]{2014prpl.conf..339T}. The dust opacity was 
calculated using Mie-theory from the optical constants of astronomical silicates \citep{weingartner_2001}. We assume that gas and dust are perfectly mixed; at $t=0$, the Stokes number (St) for a particle size of $1\,$mm is $\sim8\times10^{-4}\,(r/{\rm AU})$, indicating a good dynamical coupling between gas and the largest dust particles considered in this work. In this calculation we have assumed a mass density for the dust grains of $1$\,g\,cm$^{-3}$. For the gas line calculations we used a simple
CO abundance model, in which the usual 10$^{-4}$ H$_2$/CO abundance ratio was assumed \citep{2009ARA&A..47..481A}, but CO molecules were removed from the disc in regions 
with A$_{\rm V}\leq1.0$ due to photodissociation \rev{\citep[e.g.][]{2014ApJ...788...59W}}. In \citetalias{2017MNRAS.466.4053J} the CO abundance was also decreased by 10$^3$ where the dust temperature 
decreased below 19\,K due to freeze out of CO on the surface of dust grains. Note however that due to the high luminosity of the two central stars, the dust temperature is above $30\,$K in the whole grid.

\begin{figure}
\center
\includegraphics[width=\columnwidth]{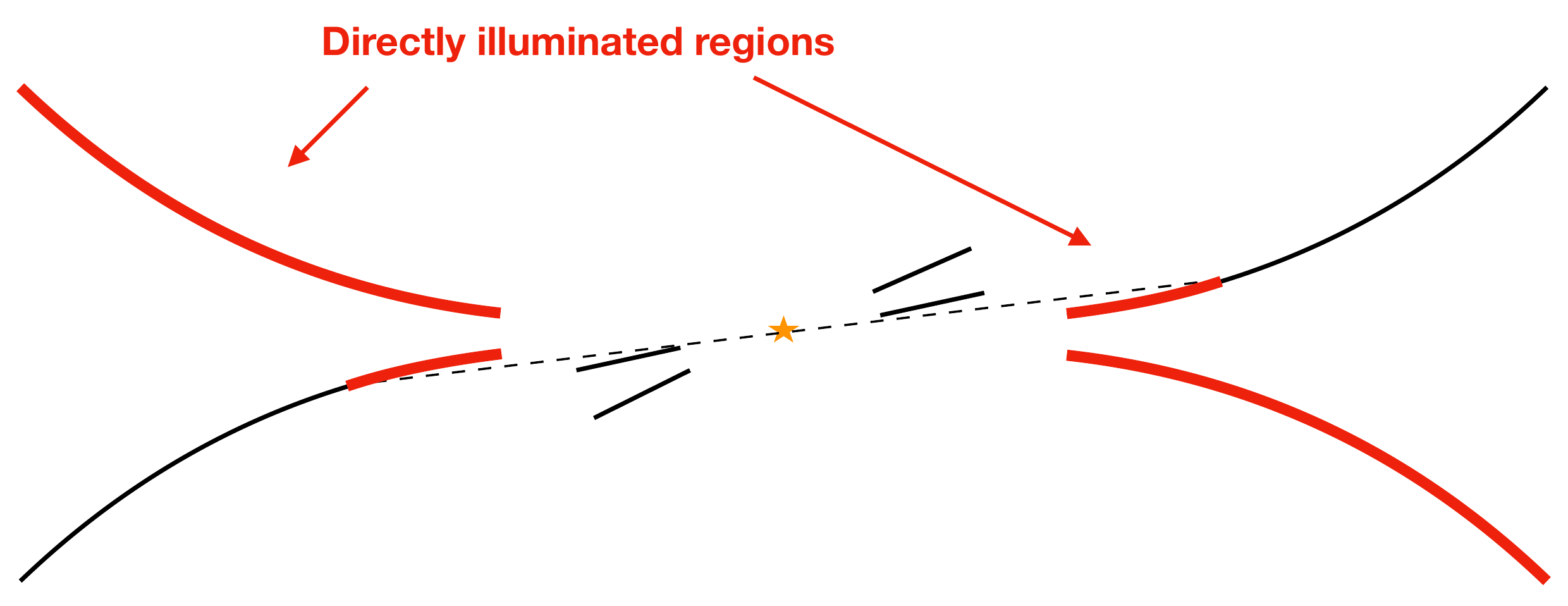}
\caption{Sketch of the shadow pattern when $\xi\gtrsim H_{\rm p}/r_{\rm break}$. While one side of the outer disc is completely illuminated by the central star, in the opposite side only the inner regions are impinged by direct radiation. Due to disc flaring, the outer regions are in the shadow of the inner disc.}
\label{fig:sketch}
\end{figure}

Once we obtained the density structure on the regular grid for the radiative transfer calculations we computed the dust temperature with \textsc{radmc-3d}. Finally we calculated continuum images at 1.65\,$\mu$m (H-band) and at 880\,$\mu$m (ALMA Band-7) as well as 
channel maps in the CO J=3-2 transition at 345.7959899\,GHz. For the CO channel maps the velocity resolution was 0.42\,km/s corresponding
to approximately 488kHz frequency resolution, the lowest channel spacing of the ALMA correlator in frequency-division mode. 
The distance of the source was taken to be 100\,pc.

To simulate observations we convolved the H-band images with a two dimensional Gaussian kernel with a full-width at half maximum (FWHM) of 
0.04\arcsec. This corresponds to the resolution of current state-of-the-art near-infrared cameras such as SPHERE/VLT \citep{2008SPIE.7014E..18B} or GPI/Gemini \citep{2008SPIE.7015E..18M}. For the sub-millimetre ALMA images we used the {\tt simobserve} and {\tt clean} tasks in Common Astronomy Software Applications (CASA) v4.7.2 to calculate synthetic observations. For the synthetic ALMA observations the full 12m Array was used in a configuration
(alma.out18.cfg) that results in a synthesised beam of 0.095\arcsec$\times$0.085\arcsec. The integration time was taken to be 4\,h for the line
 and 0.5\,h for the continuum observations, respectively. The source declination was taken to be $\delta$=-25\degr and the observations were symmetric 
 to transit. For the continuum simulations we used the full 7.5\,GHz bandwidth of the ALMA correlator in time-division mode. No thermal noise was added to any of the synthetic observations.

\section{Observational diagnostics}
\label{sec:results}

We computed synthetic images of many different snapshots of the hydrodynamical simulation. The main quantity that is going to determine a difference appearance is the misalignment angle $\xi$ between inner and outer disc. We will discuss synthetic observations for two cases, with $\xi\sim74\degr$ and $30\degr$, respectively at $t=430\,T_{\rm b}$ and $t=270\,T_{\rm b}$, in order to show results with two very different misalignment angles.

Since the disc structure is genuinely 3D, other two angles define the synthetic observations: the inclination of the outer disc in the plane of the sky $i$, and the azimuth angle around the rotation axis of the disc $\phi$. Results will be shown for three different inclination angles ($i=0\degr$, $45\degr$ and $90\degr$) and 4 different azimuth angles ($\phi_0=0\degr$, $90\degr$, $180\degr$ and $270\degr$), as in \citetalias{2017MNRAS.466.4053J}. 

\subsection{Scattered light}

\label{sec:scat}

The continuum emission from the surface layers of the disc at $1.65\,\mu$m is dominated by scattered light. The maximum temperature in the disc from the radiative transfer is $\sim400\,$K at the tidal truncation radius, thus the thermal emission in the $H$-band is negligible. The images obtained for the two misalignment angles are reported in \autoref{fig:scat_1} and \autoref{fig:scat_2}. In both cases, the image is shown with a mask mimicking a coronograph centered on the central stars, with a radius of $0.1\arcsec$.

\begin{figure*}
\begin{center}
\includegraphics[width=\textwidth]{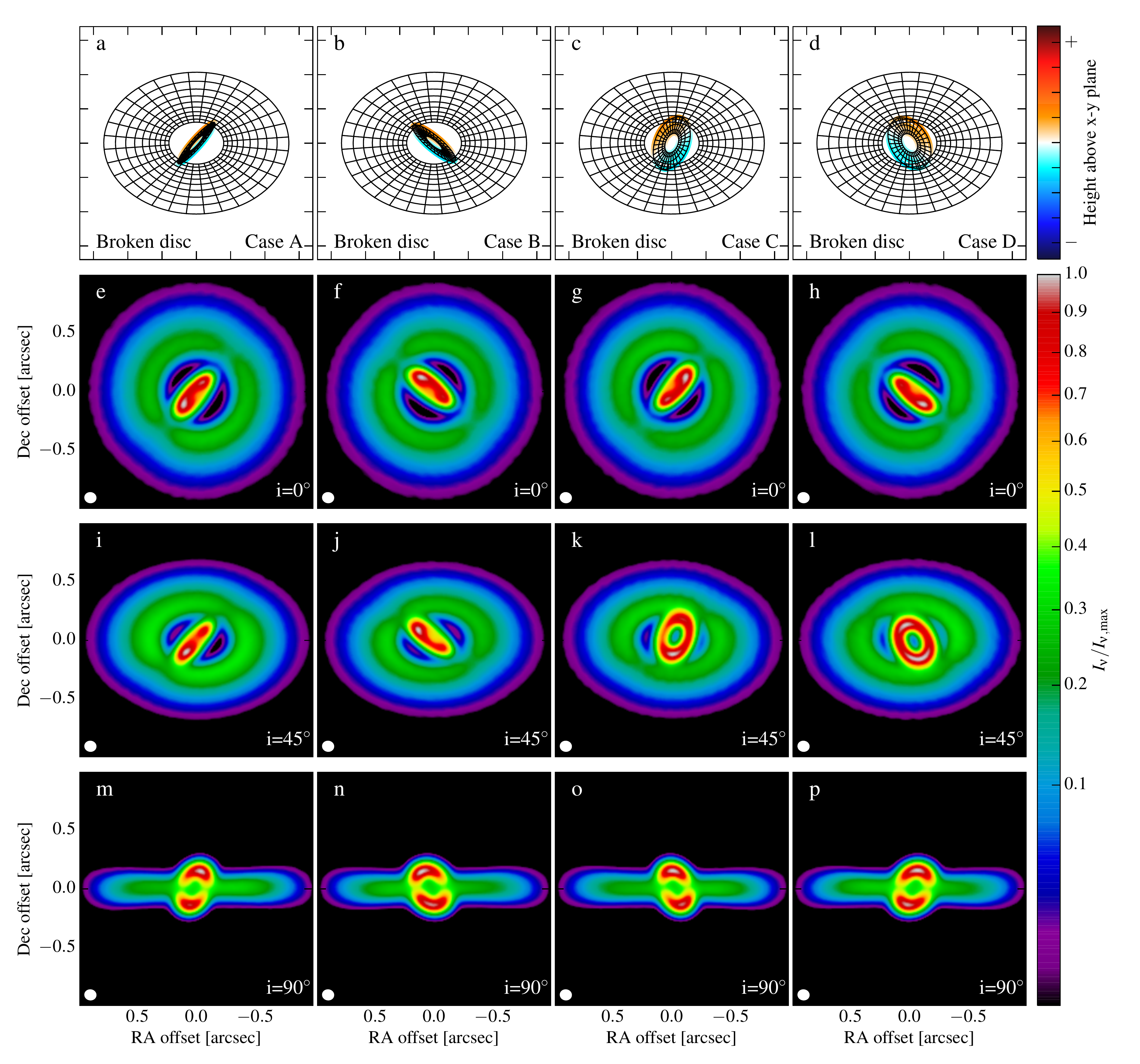}
\end{center}
\caption{Synthetic ALMA observations at $870\,\mu$m (Band 7) of the hydro model at $t=430\,T_{\rm b}$, when the misalignment angle between inner and outer disc is $\xi\sim74\degr$ (as in \autoref{fig:scat_1}). The synthesised beam is shown in white in the bottom left corner of each panel. The azimuthal regions where scattered light observations show a shadow in the outer disc present lower brightness that the directly illuminated areas, indicating that the dust mid-plane temperatures are lower.
}
\label{fig:sub_1}
\end{figure*}

\begin{figure*}
\begin{center}
\includegraphics[width=\textwidth]{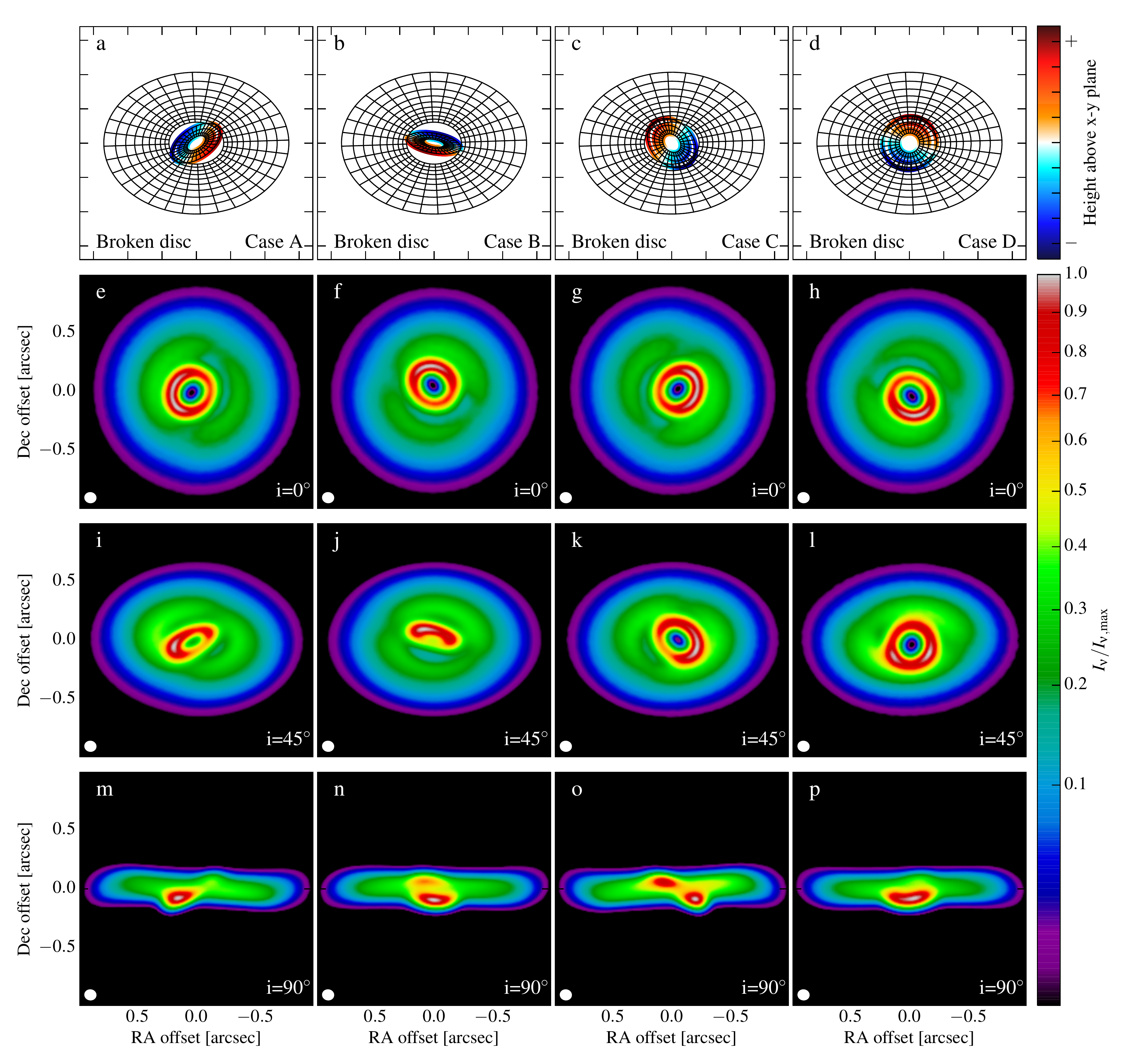}
\end{center}
\caption{Synthetic ALMA observations at $870\,\mu$m (Band 7) of the hydro model at $t=270\,T_{\rm b}$, when the misalignment angle between inner and outer disc is $\xi\sim30\degr$ (as in \autoref{fig:scat_2}). The $i=0\degr$ inclination cases show mild azimuthal asymmetries in the outer disc, with a contrast as high as $\sim5\%$ between the shadowed and illuminated sides. 
}
\label{fig:sub_2}
\end{figure*}

In the first case (\autoref{fig:scat_1}), when the inner and outer discs are mutually misaligned by $\xi\sim74\degr$, the inner disc casts two well-defined azimuthal shadows onto the outer disc, since the inner disc is optically thick at these short wavelengths. Observationally, these two shadows have been used to determine the inclination and position angle of the inner disc. In particular, the displacement of the line connecting the two shadows from the central star, and the position angle of the shadow, can be used to constrain both the inclination and the position angle of the inner disc, provided that the inclination of the outer disc is known \citep{2017A&A...604L..10M}. In all panels, the inner disc is marginally visible at the edge of the coronograph. In the panels i-j, the back side of the disc can also be seen in the southern part of the images.

\autoref{fig:scat_2} shows instead the simulated scattered light observations for an inner-outer disc misalignment angle of $\sim30\degr$. In this case, the images look quite different from the higher misalignment angle. The face-on and $45\degr$ inclination images show a strong azimuthal asymmetry, with almost half of the outer regions of the disc being in the shadow casted by the inner disc. The inner regions of the outer disc still show a double-shadow. This signature is more apparent in the right panel of \autoref{fig:scat_comp}, where we show the same image of \autoref{fig:scat_2}h, but with the emission multiplied by $R^2$ (with $R$ being the radial coordinate in the plane of the sky) to compensate for the geometric dilution of the stellar flux. This double shadow is expected to arise whenever the inner disc is misaligned by $\gtrsim 2H/R$\,rad ($\sim5\degr$ with the parameters of our simulation), where in this simple estimate we have assumed that the $\tau=1$ surface in the NIR is at $\sim H/R$. The actual surface is expected to lie at higher layers, but the exact number depends on many quite unconstrained parameters, as the grain size distribution and vertical settling. The shadows in the inner regions of the outer disc are azimuthally more extended that in the previous case, which is due to the lower misalignment angle $\xi$. The minimum value that the azimuthal width of the shadows can reach is $\sim (H/R)_{\rm in}$\,rad, i.e. the scaleheight of the inner disc, when the misalignment angle is $90\degr$. The azimuthal width will increase for lower misalignments \citep{2017ApJ...838...62L}. Combining this information with the inclination and position angle of the inner disc derived as mentioned above \citep{2017A&A...604L..10M}, it is observationally possible to put stringent constraints on the inclination, position angle, and scaleheight (of the NIR $\tau=1$ surface) of the inner disc.

A second effect that is clearly visible both in \autoref{fig:scat_2} and in the right panel of \autoref{fig:scat_comp} is the impact of flaring. Between the two shadows cast by the inner disc (in the west side of the disc), the inner region of the outer disc are still illuminated, whereas the outer regions are not. Between the inner regions of the outer disc (at $\sim28$\,AU, i.e. $r_{\rm break}$ rescaled to physical units) and the very outer regions (at $100\,$AU), the aspect ratio of the disc varies by $60\%$. If the $\tau=1$ surface of the inner disc lies between these two angles, the outer disc is going to be illuminated only in the inner regions, whereas the outer ones will be in the shadow (as portrayed in the sketch of \autoref{fig:sketch}). A flat disc would not show such a behaviour, since there would not be any radial gradient in the aspect ratio.

\begin{figure*}
\begin{center}
\includegraphics[width=\textwidth]{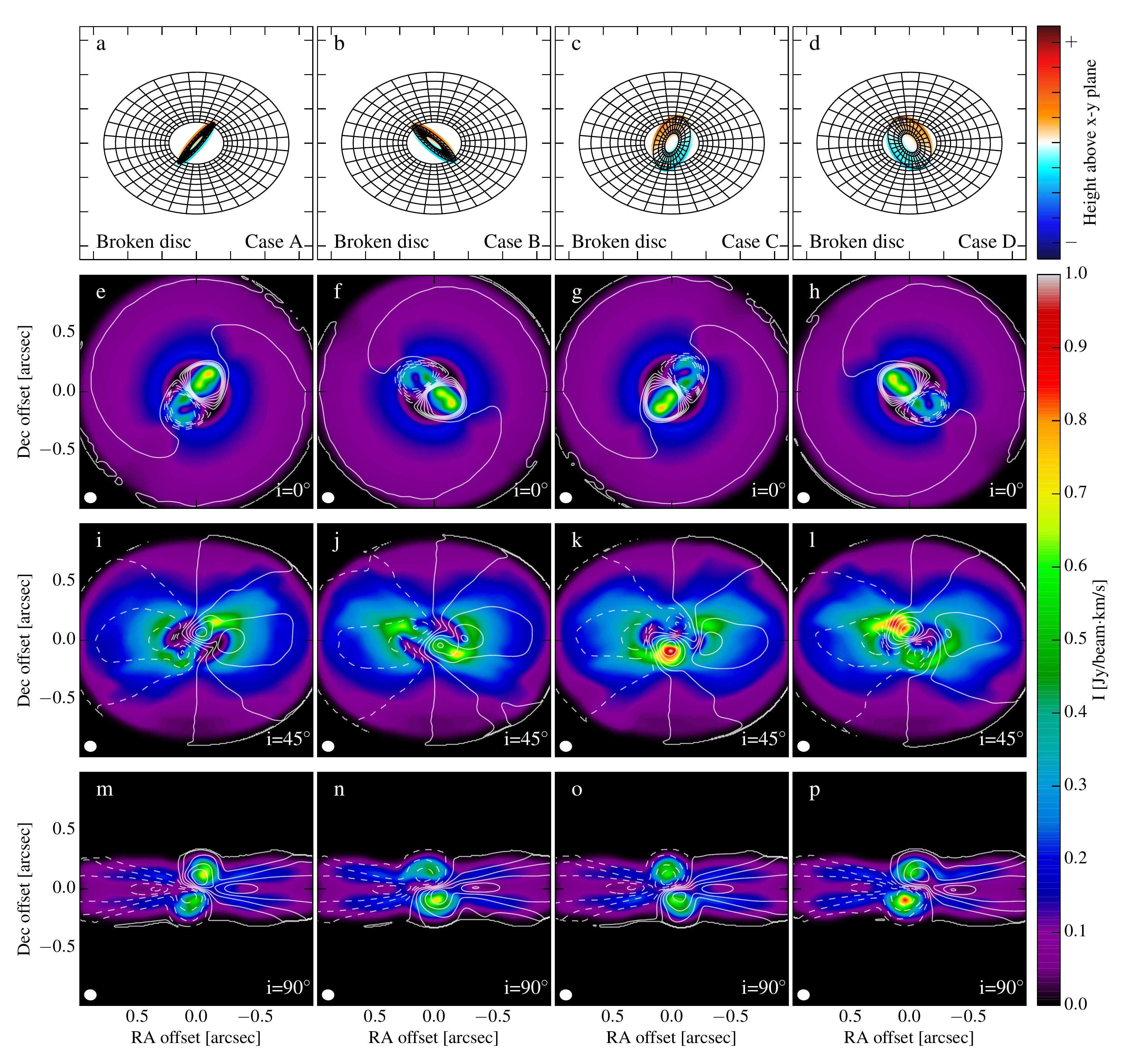}
\end{center}
\caption{Synthetic ALMA integrated intensity (0th moment) maps of CO $J$=3-2 for the $\xi=74\degr$ misalignment case. Contours indicate the intensity weighted velocity (1st moment) maps, where dashed contours denote negative velocities. For the $i=0\degr$ cases, contours are 7 evenly spaced between $-3$ and $+3\,$km\,s$^{-1}$. For both $i=45\degr$ and $i=90\degr$, contours range between $-9$ and $+9\,$km\,s$^{-1}$.
}
\label{fig:mom0_1}
\end{figure*}

\begin{figure*}
\begin{center}
\includegraphics[width=\textwidth]{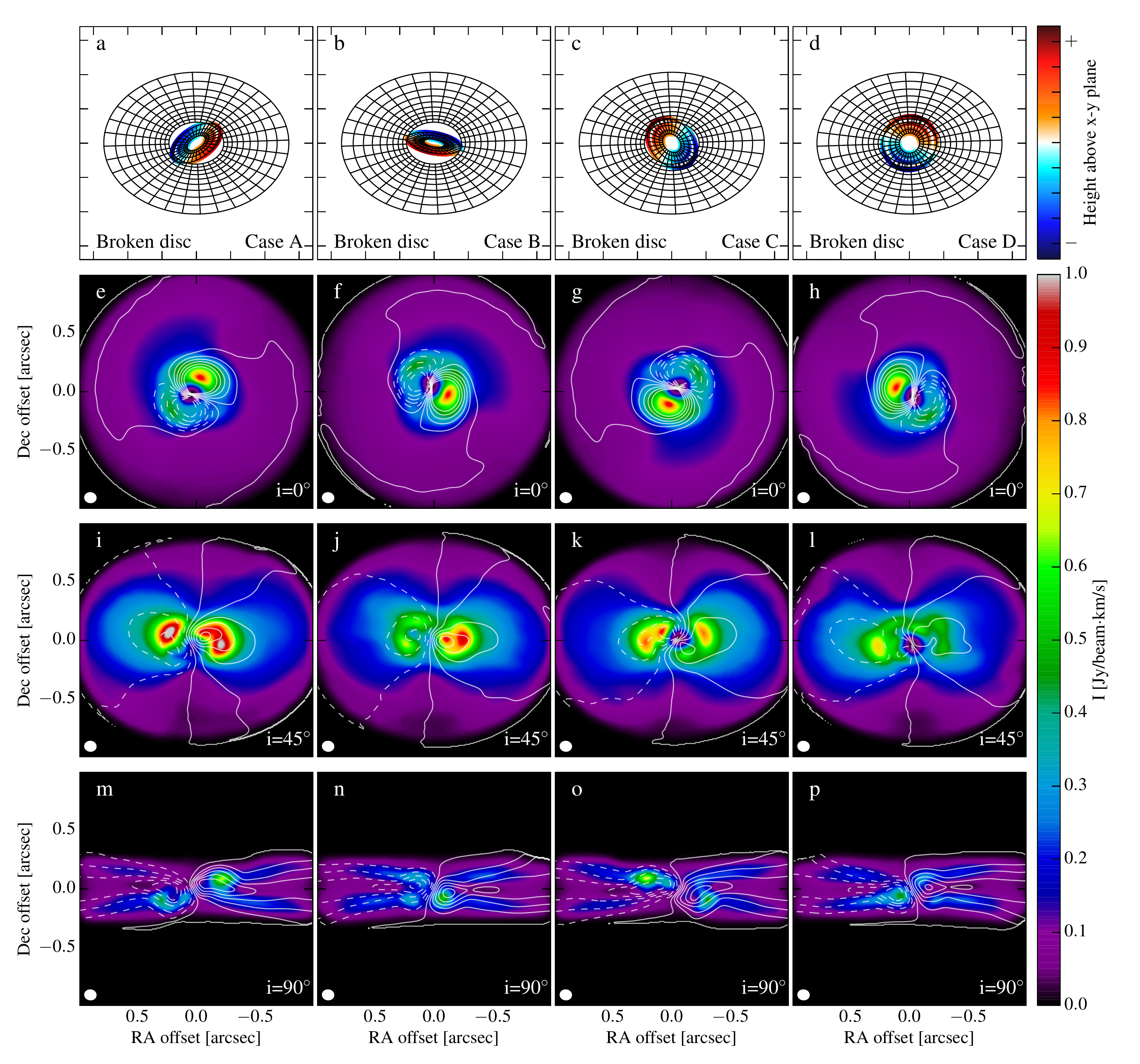}
\end{center}
\caption{Synthetic ALMA integrated intensity (0th moment) maps of CO $J$=3-2 for the $\xi=30\degr$ misalignment case. Contours indicate the intensity weighted velocity (1st moment) maps, where dashed contours denote negative velocities. Contour levels are as in \autoref{fig:mom0_1}.
}
\label{fig:mom0_2}
\end{figure*}

\subsection{(Sub-)mm continuum}

\label{sec:mm}

The (sub-)mm continuum images at $870\,\mu$m for the two inner-outer disc misalignment angles are shown in \autoref{fig:sub_1} and \autoref{fig:sub_2}. The antenna configuration used to produce the synthetic images provides high enough angular resolution to resolve both the inner and the outer disc. Whereas scattered light images show the illumination pattern of the upper layers of the disc, the surface brightness of the (sub-)mm continuum probes the colder disc midplane, where the emission is dominated by a combination of the dust column density and temperature. Panels e-h of \autoref{fig:sub_1} show maps of the $74\degr$ misalignment case with the outer disc being face-on. As expected, the inner disc is seen almost edge-on. The inner regions of the outer disc (outside $r_{\rm break}$) show a non axisymmetric structure, with two azimuthal regions of lower surface brightness, which coincide with the two shadows cast by the inner disc onto the outer regions. This is a direct effect of the dust temperature being lower in the shadowed regions (the column density is close to being azimuthally symmetric in the outer disc), as discussed in \citetalias{2017MNRAS.466.4053J}. The difference in surface brightness is $\sim20\%$ between directly illuminated and shadowed regions. The same pattern can be seen in the $45\degr$ inclination cases. Interestingly a similar temperature structure has been inferred for the HD142527 disc \citep{2015ApJ...812..126C}, and coincides with the shadows seen in scattered light \citep{2015ApJ...798L..44M}

The same effect is also observed in the synthetic maps of the $\xi=30\degr$ case (\autoref{fig:sub_2}). Here the contrast in surface brightness between illuminated and shadowed regions is however lower than in the $\xi=74\degr$ case, at about the $5\%$ level in the face-on cases.

\subsection{Moment maps of CO lines}
\label{sec:co}

\begin{figure*}
\begin{center}
\includegraphics[width=\textwidth]{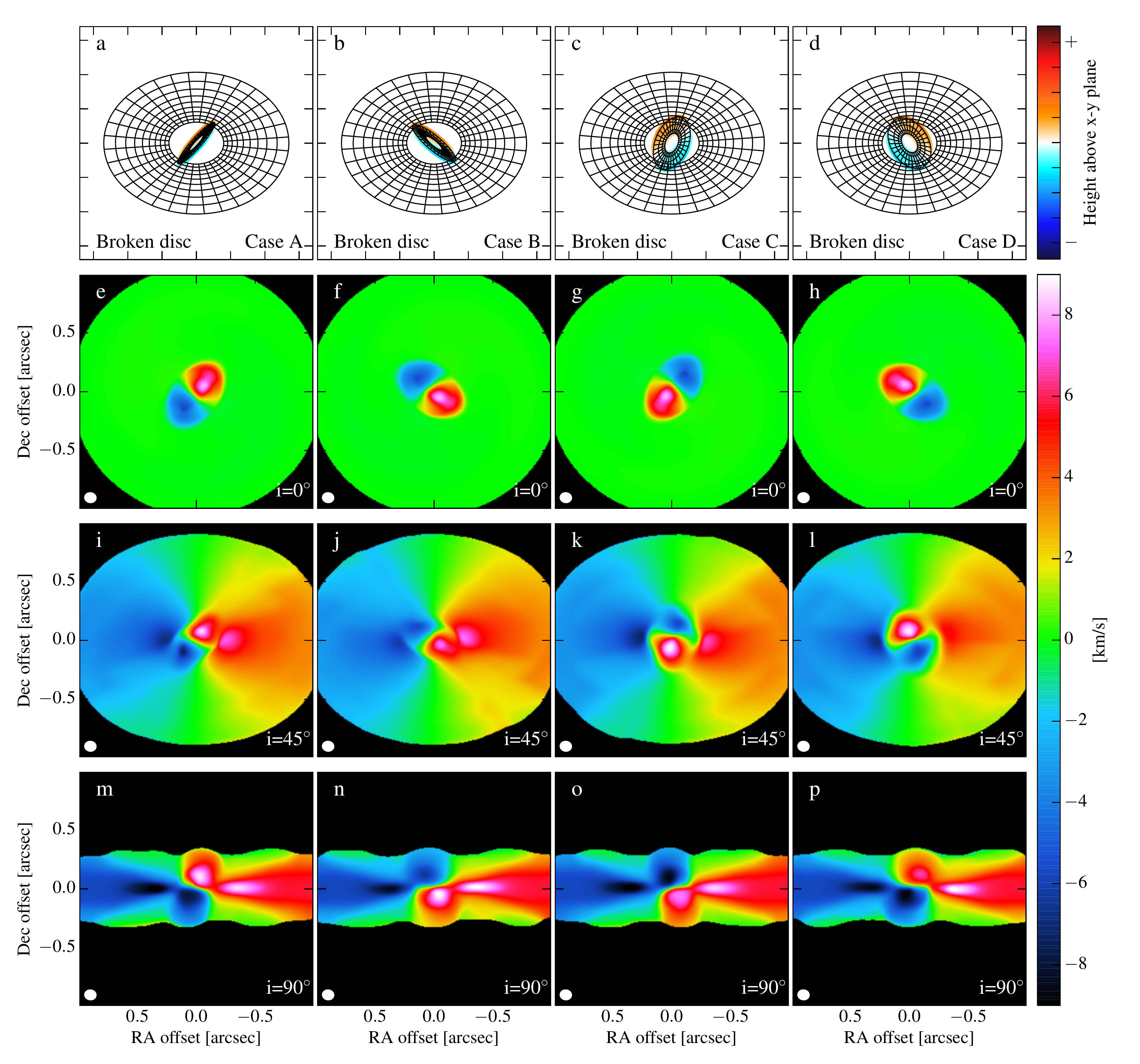}
\end{center}
\caption{Synthetic ALMA intensity weighted velocity (1st moment) maps of CO $J$=3-2 for the $\xi=74\degr$ misalignment case. Panels j and k show an inner disc that is apparently counter-rotating with respect to the outer disc, due to projection effects. As soon as the outer disc in inclined in the plane of the sky, a twist is readily observed in the 1st moment maps.
}
\label{fig:mom1_1}
\end{figure*}

\begin{figure*}
\begin{center}
\includegraphics[width=\textwidth]{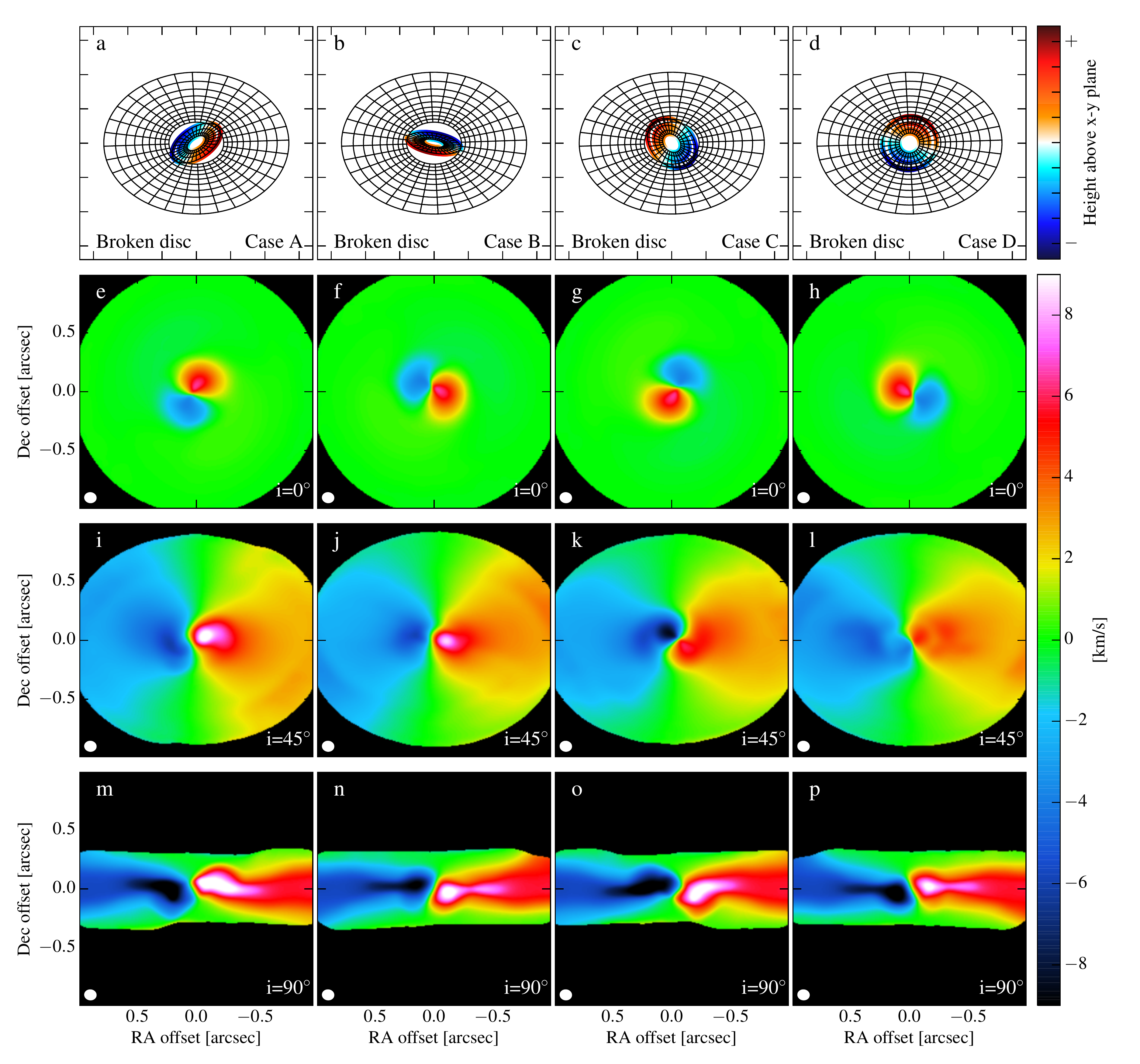}
\end{center}
\caption{Synthetic ALMA intensity weighted velocity (1st moment) maps of CO $J$=3-2 for the $\xi=30\degr$ misalignment case. When the outer disc is inclined in the plane of the sky, a twist in the 1st moment map is observed whenever the line of nodes is not along the semi-major axis of the outer disc (panels j and l). When the line of nodes is parallel to the semi-major axis, the inner disc shows either higher (panel i) or lower (panel k) velocities than if the inner disc were rotating on the same plane of the outer disc.
}
\label{fig:mom1_2}
\end{figure*}

The synthetic ALMA images of the CO $J$=3-2 line are shown in \autoref{fig:mom0_1} - \autoref{fig:mom1_2}, where the first two figures portray the 0th moment maps, and the second two figures show the 1st moment maps. As in \autoref{sec:scat} - \autoref{sec:mm}, both the high and low misalignment angle cases are reported. Spatially resolved line emission provides information both on the surface brightness and on the gas kinematics, which is a key diagnostics of warped structures.

As shown in \citetalias{2017MNRAS.466.4053J}, CO lines are much more sensitive to the illumination pattern than the mm continuum. In particular, rotational transitions of CO become optically thick at very low column densities, as opposed to the optically thinner continuum emission at similar wavelengths. Thus, while the latter provides information on the disc midplane, the former traces the thermal conditions of the upper layers of the disc, where the line originates. Being CO in local thermo-dynamical equilibrium (LTE) in its rotational ladder for typical conditions of protoplanetary discs, the emission is mostly sensitive to the temperature at the $\tau=1$ layer of the emission line \citep[e.g.][]{2013A&A...559A..46B,2017A&A...605A..16F}. This is clearly shown by comparing the ALMA integrated intensity maps (\autoref{fig:mom0_1}-\autoref{fig:mom0_2}) with the scattered light images (\autoref{fig:scat_1}-\autoref{fig:scat_2}), in particular the face-on cases, panels e-h. The pattern of the shadow cast by the inner disc as shown by scattered light is well reflected in the surface brightness (thus in the temperature structure) of the CO line. While the contrast in brightness between the directly illuminated and shadowed regions is as high as $\gtrsim100$ in the scattered light images, it decreases to a factor of $\sim2$ in the CO integrated intensity maps.

In panels e-l ($i=0\degr$ and $i=45\degr$) in both \autoref{fig:mom0_1} and \autoref{fig:mom0_2} the cavity carved by the binary within the inner disc is observable at the chosen resolution. In the $\xi=74\degr$ case (\autoref{fig:mom0_1}) the disconnection between inner and outer disc is well visible as well, in particular in the cases where the inner disc is seen almost edge on. The gap between the inner and outer disc, which is due to a discontinuity in the angle $\beta$, cannot be observed if the disc is not actually broken, but only warped. This can be easily noticed by comparing \autoref{fig:mom0_1} with figure 9 of \citetalias{2017MNRAS.466.4053J}, where was shown the total intensity map of the same line for a warped disc. The azimuthal variation in surface brightness is also much more severe in the broken disc case, where the misalignment angle between inner and outer disc can be $\gg H_{\rm p}/R$.

Together with the scattered light pattern, the best provider of information about the 3D structure of the disc is the intensity weighted velocity (1st moment) map. The velocity structure of a system with a tilted inner disc is significantly different from a simpler co-planar case. This can be clearly seen in both \autoref{fig:mom1_1} and \autoref{fig:mom1_2}, where the first one portrays 1st moment maps of the $\xi=74\degr$ case, and the second one of the $\xi=30\degr$ case. Panels e-h in both figures show an outer disc that is seen face-on. As the inner disc is tilted, the rotation pattern stands out in these inner regions, with a magnitude of the projected velocity that is much higher than if the inner disc were face-on. As the zero projected velocity line runs along the semi-minor axis of a disc, the position angle of the inner disc can be easily derived by looking at this line, as it is routinely done for full co-planar discs. By comparing the cartoons in panels a-d with the 1st moment map of the face-on case in panels e-h, it is clear that the zero projected velocity line is perpendicular to the line of nodes, i.e. the line that defines the interception between the planes of the inner and the outer disc.

As the outer disc becomes more inclined in the plane of the sky, the 1st moment map becomes more complicated. At $i=45\degr$, \autoref{fig:mom1_1} shows an intensity weighted velocity map that is clearly significantly twisted for all azimuth angles (panels i-l). The twisting is much more significant than the warp cases addressed in \citetalias{2017MNRAS.466.4053J}, since the misalignment angle between inner and outer disc is much larger. The different pattern between panels i and j is due to the $90\degr$ difference in the position angle of the inner disc between the two cases. Even more peculiar are panels k and l, where the inclination in the plane of the sky is such that the inner disc seems to be counter-rotating, whereas the actual misalignment angle between inner and outer disc is $74\degr$. While in this case the counter-rotation is just an effect of velocity projection onto the sky, note that as the inner disc precesses, counter-rotation can actually be attained in the hydrodynamical simulation (see \autoref{fig:xi}), where with counter-rotation we consider snapshots where $\xi>90\degr$.

As $\xi$ becomes lower, the twisting in the 1st moment maps become less severe. In panels i and k of \autoref{fig:mom1_2}, where $\xi=30\degr$, the very inner regions of the disc still show a quite significant twist. Panels j and j, instead, show a very minor twist. The reason is simply that in this second case the line of nodes is close to be along the semi-major axis of the outer disc, thus the zero velocity map does not bend significantly in the intensity weighted velocity map. However, there is a clear difference between panel j and panel l: the inner regions of the disc of the former show projected velocities that are higher by $80\%$. This can be simply understood by geometrical means: panel i has an inner disc where the far side is tilted away from the observer, whereas panel k has an inner disc where the far side is tilted towards the observer. Due to projection effects, the former shows projected velocities that are higher than if the disc were entirely co-planar, whereas the latter presents lower velocities compared to the co-planar case \citep[see also][\citetalias{2017MNRAS.466.4053J}]{2012ApJ...757..129R}. As discussed in \citet{2014MNRAS.442.3700F} and \citetalias{2017MNRAS.466.4053J}, this effect is maximised for very low inclination angles $i$ of the outer disc.

In the estimates of the gas temperature (and dust temperature in the disc midplane in \autoref{sec:mm}), we have made the implicit assumption that the gas reacts instantaneously to the local illumination, i.e. to the local light spectrum. This assumption is equivalent to assuming a cooling timescale $t_{\rm cool}$ that is much shorter than the local dynamical timescale $1/\Omega$, where $\Omega$ is the orbital frequency. Being more precise, the gas temperature can react to the azimuthal variations of the shadow pattern if $(\Delta \phi/2\pi)/\Omega > t_{\rm cool}$, where $\Delta\phi$ is the azimuthal width of the shadow in radians, i.e. if the time for a gas parcel to cross the shadowed region is longer than the cooling timescale. This assumption is probably valid for the upper layers of the disc, where the CO emission generates, since both the heating and cooling timescales are rather fast. However, this might not hold true for the disc midplane \citep[e.g.][]{2016ApJ...823L...8M}, where the cooling timescale is much longer; thus the significance of the azimuthal asymmetries predicted for the (sub-)mm continuum are a reasonable upper limit to what is expected in real systems.

\section{Discussion}

We have shown the observational variety that can arise from a protoplanetary disc that presents a tilted inner disc. The hydrodynamical simulation presented in this paper are specific for a circumbinary disc that is misaligned with respect to the orbital plane of the central equal-mass binary. Published observations of particular protoplanetary discs are showing that there are other means to break a disc. For example, in HD142527, the inner disc \citep{2015ApJ...798L..44M} is likely to be misaligned due to the interaction with the $\sim0.2M_\odot$ companion \citep{2012ApJ...753L..38B,2014ApJ...781L..30C} orbiting within the transition disc cavity, but outside the circumprimary disc. A different case is AA Tau, where models of the optical and NIR light-curve \citep[e.g.][]{2013A&A...557A..77B} and new ALMA images \citep{2017ApJ...840...23L} of an almost edge-on inner disc invoke an interaction between the inner disc and the magnetic dipole moment of the central star \citep{2011MNRAS.412.2799F}. However, the qualitative properties of the observational diagnostics presented in \autoref{sec:results} can be extended to these different scenarios, as long as a misaligned inner disc casts a shadow onto the outer regions.

In some transition discs with large dust cavities, the interaction with a potentially misaligned companion has successfully modelled many azimuthal asymmetries observed in the ring of the (sub-)mm continuum \citep[e.g.][]{2017MNRAS.464.1449R}. This might indicate that indeed misaligned companions might be at the origin of the central cavity \citep[e.g.][]{2012A&A...545A..81P}, leading to the azimuthal asymmetries in the outer dust trap, and tilting the inner disc as in HD142527 system \citep{2017MNRAS.469.2834O}. However, conclusive proof will come only when these potential companions will actually be observed.

In the synthetic scattered light observations, no sign of a spiral-like pattern is observed. In order for the inner disc to illuminate the outer disc with a spiral morphology, a strong twisting is expected to arise in the inner disc. However, in \autoref{fig:beta_gamma_r} we demonstrate that the inner disc is not twisted, at least for the particular case of an inner binary potential. In \citetalias{2017MNRAS.466.4053J}, we have also shown that even when the disc does not break, since the misalignment angle between the binary and the disc is not high enough, the disc is not significantly twisted for typical protoplanetary conditions. If the precession timescale of the inner disc is comparable to the light travel time between the stars and the outer disc, then a spiral pattern in the shadows of the upper layers of discs might be observed \citep{2016A&A...593L..20K}. In this paper we have not taken this effect into account, but we stress that in order to have a precession timescale that is of the order of the photon travel time through the disc, a rather compact binary is needed (with $a\sim0.1\,$AU). Finally, the potential triggering of spiral waves due to the asymmetric shadowing, generating azimuthal asymmetries in the pressure profile \citep{2016ApJ...823L...8M} has not been considered here. A more detailed discussion on the gas and dust temperatures resulting from the asymmetric illumination is included in \autoref{sec:co}.

The particular twisted 1st moment maps shown in \autoref{sec:co} have already been observed in a few sources by ALMA \citep[e.g.][]{2014ApJ...782...62R,2015ApJ...811...92C,2017A&A...597A..32V,2017ApJ...840...23L,2017arXiv171000703W}, in particular within the cavities of transition discs. In most cases the angular resolution was not high enough to resolve the twist, but a thorough analysis has shown that the inner regions of the disc have velocities that are higher than expected from a co-planar model. \citet{2014ApJ...782...62R} realised that gas radial inflows can lead to the same twisted velocity maps. Theoretical models have struggled to explain why such inflows should be present in systems hosting a single star. However, recent studies on magnetised winds are starting to show that these inflows might be a consequence of fast accretion driven by the winds \citep{2017ApJ...835...59W,2017arXiv170104627Z}. To observational distinguish between radial inflows and warped or tilted inner discs, scattered light observations can play a fundamental role. Whereas the projected velocity map can be degenerate between the two cases, the azimuthal asymmetries shown in \autoref{sec:scat} can arise only if an actual misalignment is present. Thus, multi-wavelength observations in the NIR and in the (sub-)mm to constrain the gas kinematics are fundamental to discriminate between the two different scenarios. Moreover, while radial inflows will always lead to higher projected velocities than what is expected from Keplerian and co-planar models, tilted inner disc can provide both higher and lower velocities (see discussion in \autoref{sec:co}), depending on whether the inner disc bends towards or away from the observer.

A warped or broken inner disc and radial inflows are not mutually exclusive in real systems. In particular, \citet{2015ApJ...811...92C} suggested that the HD142527 system might host both physical mechanisms, with inflowing gas originating from the angular momentum cancellation of neighboring precessing annuli. As shown in \autoref{fig:accretion}, this might occur in circumbinary discs. However, in our simulations the infalling material is mostly confined between the inner and the outer disc, and the observability of the gas inflow would be extremely challenging due to the high angular resolution it would require. In HD142527, the binary is not of equal mass and the orbit of the secondary star is most likely eccentric \citep{2016A&A...590A..90L}, which can lead to a much more significant gas inflow compared to our circular binary case. Moreover, in HD142527 it is the circumprimary disc that is misaligned with respect to the circumbinary disc. The radial inflow between the circumprimary and circumbinary discs in HD142527 is on a much larger spatial scale, and thus easier to observe, compared to the very narrow region between the two parts of the circumbinary disc in our models.

\section{Summary and conclusions}

In this paper, we have modeled a misaligned circumbinary disc, where the initial misalignment between the disc and the central binary leads to the formation of a tilted inner disc precessing as a rigid body. We have post-processed snapshots of the hydrodynamical simulation at different stages of the evolution, in order to provide observational diagnostics of a tilted inner disc at different wavelengths, motivated by actual hydrodynamical models. Our main conclusions can be listed as follows:

\begin{enumerate}
\item The misalignment angle between inner and outer discs is a function of both the initial misalignment with respect to the central binary, and of the precession phase of the inner disc; in our simulation, the misalignment angle can become as large as $\gtrsim90\degr$.
\item We confirm earlier studies showing that shocks can occur where the inner disc grazes the outer one, leading to significant angular momentum cancellation and consequent accretion.
\item Synthetic scattered light observations show strong azimuthal asymmetries; their pattern depends strongly on the instantaneous misalignment between inner and outer disc.
\item The asymmetric illumination of the disc leads to surface brightness variations both in the disc midplane, traced by (sub-)mm continuum, and even more significantly in the disc upper layers, traced by optically thick lines, e.g. from CO rotational transitions.
\item The intensity weighted velocity maps can readily indicate the presence of a broken inner disc, whenever the inner disc is resolved.
\item A combination of scattered light observations to determine the 3D geometry of the inner disc, and line observations to determine the disc kinematics, are a powerful tool to discriminate between radial inflows and warped/tilted structures.
\end{enumerate}


\section*{Acknowledgements}

We are grateful to the referee, Chris Nixon, for his thorough reading of the manuscript, and his comments which helped improving the clarity of the paper. We are thankful to Myriam Benisty for fruitful discussions. SF is thankful to E.~F. van Dishoeck for her support. This work has been supported by the DISCSIM project, grant agreement 341137 funded by the European Research Council under ERC-2013-ADG. This research was supported by the Munich Institute for Astro- and Particle Physics (MIAPP) of the DFG cluster of excellence ``Origin and Structure of the Universe''. \autoref{fig:evolution} and \autoref{fig:accretion} were produced using \textsc{splash} \citep{2007PASA...24..159P}, a visualisation tool for SPH data. Most of the other figures were generated with the \textsc{python}-based package \textsc{matplotlib} \citep{2007CSE.....9...90H}.

\bibliographystyle{mnras}
\bibliography{bib_warps}




\bsp	
\label{lastpage}
\end{document}